\documentclass[twocolumn]{aastex631}

\newcommand{\e}{$^{-1}$}

\newcommand{\eee}{$^{-3}$}
\newcommand{\msun}{\mathrm{M}_\odot}
\newcommand{\alphaturb}{\alpha_{\mathrm{vir}}}
\newcommand{\alphaB}{\alpha_{\mathrm{B}}}

\newcommand{\tff}{t_\mathrm{ff}}
\newcommand{\SFE}{\mathrm{SFE}}
\newcommand{\hi}{\texttt{M2e4\_mu0.4}}
\newcommand{\med}{\texttt{M2e4}}
\newcommand{\lo}{\texttt{M2e4\_mu4.2}}
\newcommand{\starforge}{{\sc starforge}}

\usepackage{amsmath} 

\begin{document}

\title{The Life and Times of Star-Forming Cores: an Analysis of Dense Gas in the STARFORGE Simulations}

\author[0000-0003-1252-9916]{Stella S. R. Offner}
\affiliation{Department of Astronomy, University of Texas at Austin \\
Austin, TX, 78712 USA}

\author{Josh Taylor}
\affiliation{Department of Astronomy, University of Texas at Austin \\
Austin, TX, 78712 USA}

\author{Michael Y. Grud\'ic}
\affiliation{ Center for Computational Astrophysics, Flatiron Institute \\
162 Fifth Avenue, New York, NY 10010, USA}



\begin{abstract}
Dense gas in molecular clouds is an important signature of ongoing and future star formation.
We identify and track dense cores in the \starforge\ simulations, following the core evolution from birth through dispersal by stellar feedback for typical Milky Way cloud conditions. Only $\sim$8\% of cores  
host protostars, and most disperse before forming stars. The median starless and protostellar core lifetimes are $\sim 0.5-0.6$\,Myr and $\sim0.8-1.1$\,Myr, respectively, where the protostellar phase lasts $\sim 0.1^{+0.1}_{-0.05}$\,Myr.
While core evolution is stochastic, we find that virial ratios and linewidths decline in prestellar cores, coincident with turbulent decay. Collapse occurs over $\sim 0.1$\,Myr, once the central density exceeds $\gtrsim 10^6$cm\eee. Starless cores, only, follow linewidth-size and mass-size relations, $\sigma \propto R^{0.3}$ and $M \propto R^1$. The core median mass, radius, and velocity dispersion scale weakly with the cloud magnetic field strength. We cluster the core properties and find that protostellar cores have $>80$\% likelihood of belonging to three particular groups that are characterized by high central densities, compact radii, and lower virial parameters. Overall, core evolution appears to be universally set by the interplay of gravity and magnetized turbulence, while stellar feedback dictates protostellar core properties and sets the protostellar phase lifetime. 

\end{abstract}

\keywords{stars:formation,ISM:kinematics and dynamics,numerical methods:MHD,turbulence}


\section{Introduction} \label{sec:intro}

Stars form in dense cores, the densest most compact regions within molecular clouds. Due to the inefficiency of star formation, dense gas comprises a relatively small fraction of clouds by mass and volume \citep{DunhamStutz2014a}. Dense cores observed in  typical Milky Way molecular clouds have sizes of $\lesssim0.1$\,pc and number densities $\gtrsim 10^4$\,cm\eee \citep{PinedaArzoumanian2023a}. As the direct precursors to the formation of individual star systems, dense cores
provide important insights into the initial conditions of star-forming gas and serve as a proxy for incipient and future star formation.

Somewhat counterintuitively, despite their low velocity dispersions and cold temperatures a significant fraction of cores appear to be gravitationally unbound and are instead likely confined by the ambient cloud pressure \citep[e.g.,][]{Maruta_2010,BarnesYonekura2011a,Pattle_2015,KirkFriesen2017a,KerrKirk2019a}.

These observations collectively raise a variety of questions: How will identified cores evolve over time? How do different populations of cores relate to one another? What initial conditions lead to future star formation? The picture is further muddied by the wide range of definitions used to identify cores \citep{Goodman_2009a,MenshchikovAndre2010a,CurrieBerry2014a}, by incompleteness \citep{SokolGutermuth2019a,OneillCosentino2021a}, and by various observational biases, which may produce false over-densities from line-of-sight projection \citep{BeaumontOffner2013a}. Moreover, the future evolution of identified dense cores is highly uncertain. Gravitational stability arguments based on the virial theorem, which point to future collapse or dispersal, are highly simplified and prone to large errors \citep{Ballesteros-ParedesGazol2006a,SinghMatzner2021a,GangulyWalch2024a}.  In addition, the observational stability estimates do not consider the influence of the magnetic field and fail to capture the impact of the turbulent environment, which could promote collapse via accretion or disperse gas due to shocks. 

Simulations and theoretical models have provided an important complement to observational surveys. For example, the origin of the core mass function may be explained by the confluence of gravity with the universal scaling relations of turbulence \citep{Padoan_1997,Hopkins_2012,GuszejnovHopkins2015a}. Simulations also suggest the mapping between the core mass function and stellar mass function may not be so direct \citep{OffnerClark2014a,SmullenKratter2020a}. 
In addition, analysis of over-densities in turbulence simulations indicate that a large fraction of observed cores actually disperse rather than become star-forming \citep{OffnerTaylor2022a}.

However, most studies of cores to date have been conducted in simulations neglecting various physics and stellar feedback processes, adopted an artificial simulation stopping time, and included relatively small samples of cores ($< 1000$). In this work we study dense cores in the \starforge\ simulations, which include more important physical processes and forms of stellar feedback. We track cores from formation until star formation is terminated by cloud dispersal. In \S2 we describe our numerical methods, approach to core identification, and statistical methods applied to track, analyze and cluster core properties. We report our results, including derived core properties, evolution, and clustering in \S3. \S4  discusses our results in the context of prior observational and theoretical work, and we conclude in \S5. 


\begin{table*}
\hspace{-2.5cm}
	\begin{tabular}{|l|ccccc|ccccc|}
	     \multicolumn{8}{c}{}\\ 
		 \multicolumn{1}{c}{\bf } & 
		 \multicolumn{5}{c}{\bf Simulation Parameters}&
		 \multicolumn{5}{c}{\bf Outcomes} \\
		\hline
		\bf Cloud label & $\mu$ &  $\beta$ & $E_B/|E_{\rm grav}|$ & $\frac{M_{\Phi}}{M_0}$ & $B_z$ ($\mu$G) &  SFE [\%] &  $t_\mathrm{disrupt}/\tff$  & $N_*$ & $N_{\rm c, tot}$ & $N_{c,*}/N_{\rm c, tot}$ [\%] \\
		\hline
		\lo  & 4.2  &  0.78 & 0.01 & 
		0.1 & 2 & $\mathrm{9\pm 0.3}$ & $\mathrm{1.6\pm 0.2}$ & 2,174 &  391,261 & 7.5 \\ 
		\hline
		{\med} (fiducial) &  1.3  
		  &  0.078 & 0.1 & 
		0.4 & 6.3 & 7 & 2.0 & 2,246 & 363,331 & 7.9\\ 
		\hline
		\hi & 0.42  
		  &  0.0078 & 1 & 
		 4 & 20 &  5 & 2.2 & 2,415 &  333,108 & 13.2 \\ 
		\hline
	\end{tabular}
        \vspace{-0.1cm}
 \caption{Summary of the cloud initial conditions, where 
all clouds have  mass $M_0 = 2 \times 10^4~\msun$, radius $R_0 = 10$ pc, and 3-$d$ velocity dispersion $\sigma_{\rm rms} = 3.2$ km/s. 
 Since these runs explicitly evolve the radiation field, the initial gas and dust temperatures are set by the assumed ISRF. The input parameters are the mass to magnetic flux ratio $\mu$, 
 plasma $\beta = P_{\rm theremal}/P_{\rm magnetic} = 2 c_s^2/v_A^2$ where $c_s$ is the sound speed and $v_A$ is the Alfv\'en speed, magnetic virial ratio, 
 and magnetic mass scale  where $M_\Phi/M_0=\sqrt{2\alphaB}$  (note that these are all defined assuming a 10~K gas temperature), and the initial magnetic field. The last four columns display the final star formation efficiency ($\SFE=M_*/M_0$), the disruption time, i.e., the cloud lifetime, normalized to the initial cloud freefall time, the final number of stars, and the total number of identified cores across all snapshots, and the fraction of protostellar cores.  The uncertainties in the SFE and disruption time given in the first row are standard variations
of three simulations that were run with different initial turbulent seeds. See \citet{GuszejnovGrudic2020a} for an overview of the dimensionless parameters and \citet{GuszejnovGrudic2022a} for detailed star formation histories. Simulation parameters, SFEs, and disruption times are reproduced from Table 1 in \citet{GuszejnovRaju2023a}.} 
 \label{tab:IC_phys}
\end{table*}

\section{Methods} \label{sec:methods}

\subsection{The \starforge\ simulations}

In this work, we analyze three magnetohydrodynamic (MHD) simulations of star-forming clouds from the \starforge\ project. The calculations are run with the Lagrangian meshless finite-mass code {\small GIZMO}\footnote{ http://www.tapir.caltech.edu/~phopkins/Site/GIZMO.html}. The code methodology and physics modules are described in detail in \citet{GrudicGuszejnov2021a}, while the simulation initial conditions, bulk properties, and star formation histories are outlined in \citet{GuszejnovGrudic2022a}. Here, we only give a brief summary of the simulations and refer the reader to these papers for a fuller description.

\subsubsection{Numerical Overview}

The \starforge\ simulations in this investigation include all main stellar feedback processes, including radiation from protostellar and stellar sources, protostellar jets, stellar winds, and supernovae. Individual sources, represented by sink particles, follow a sub-grid prescription that tracks protostellar and main-sequence evolution as a function of source mass, accretion history, and age; these source properties in turn set the mass-loss rate, luminosity and lifetime.

The simulations evolve the magnetic field assuming ideal MHD \citep{hopkins_gizmo_mhd}. Consequently, magnetic braking is very efficient, and protostellar disks are not resolved. The gas and dust temperatures are computed using the radiative cooling and thermochemistry model described in \citet{fire3}, which includes recombination, thermal bremsstrahlung, molecular lines, metal lines, fine structure and dust collisions. The simulations co-evolve the gas, dust, and radiation temperature self-consistently using multi-band radiation transport where absorption and emission is dictated by dust opacities. In addition to internal stellar sources, the contribution of the interstellar radiation field (ISRF) is modeled by fixing the radiation field at the simulation boundary.

Each cloud is evolved  until it is dispersed by stellar feedback and nearly all star formation is quenched, which occurs after $\sim$10\,Myr. Thus, star formation concludes self-consistently when no high-density gas remains. This is a critical feature of our study, which aims to track the full life-cycle of dense cores as influenced by stellar feedback and the cloud environment.

\subsubsection{Initial conditions and cloud properties}

The simulations each model a cloud of mass $M_0=2\times 10^4\,\msun$ with a mass resolution of $\Delta m=10^{-3}\,\msun$ and radius $R=10$\,pc.  The cloud is initialized as a sphere with uniform density, $\rho_0$, and uniform magnetic field, $B_z$. The initial cloud gas and dust temperatures are set by the ISRF, set to model Solar neighborhood conditions \citep{1983A&A...128..212M}. The cloud begins near thermal pressure equilibrium with the ambient gas, which has $\rho=\rho_0/100$. The turbulence is initialized by applying a random velocity field with power spectrum $E_k \propto k^{-2}$ and amplitude set by the desired {\rm kinetic} virial parameter, $\alphaturb \equiv \frac{5 \sigma_{\rm rms}^2 R_0}{3GM_0}=2$ \citep{BertoldiMcKee1992a}, which corresponds to a velocity dispersion $\sigma_{\rm rms} = 3.2$ km/s.

 At the adopted simulation resolution, sink particles, henceforth referred to as protostars, are inserted when the gas density exceeds $3 \times 10^{-14}$ g cm\eee ($n_{\rm H} \sim 10^{10}$ cm\eee) and when the Jeans and tidal criteria are also satisfied \citep[see][for details]{GrudicGuszejnov2021a}. The freefall time for gas at this critical density is $\tff = \sqrt{3 \pi/ (32 G \rho)} \simeq 400$ years, significantly smaller than the output snapshot interval, which is $\Delta t \simeq 24,000$ years. Consequently, the protostellar core lifetime, which is measured from the first snapshot the core is identified as protostellar, is insensitive to our sink formation threshold.

We explore the impact of the cloud magnetic field on core properties by varying the initial cloud magnetic energy by a factor of 100, from 1\% to 100\% relative to the gravitational energy. The magnitude of the initial magnetic field is set by the mass-to-magnetic flux ratio, $\mu = 0.4 \sqrt{-E_{\rm grav}/ E_{\rm B}}$ \citep{MouschoviasSpitzer1976a}. We set $\mu = \{4.2, 1.3, 0.42 \}$, which corresponds to magnetic virial ratios  $\alphaB = E_B/|E_{\rm grav}| = \{0.01, 0.1, 1\}$, and adopt the intermediate field case, $\mu=1.3$, as the fiducial run.  All runs assume an initial solar metallicity ($Z=1$).
Table \ref{tab:IC_phys} summarizes the simulation parameters.

\subsection{Dense Core Analysis}

In this section we describe our methods to identify, track over time, and characterize dense cores in the simulations.

\subsubsection{Core Identification}\label{sec:identification}

We use  {\it astrodendro}\footnote{\url{http://dendograms.org}}, an open-source {\it Python} package    to identify cores in each simulation snapshot \citep{Rosolowsky_2008b,Goodman_2009a}.  The dendrogram algorithm is a essentially a watershed algorithm that first finds the local peaks above some threshold and level of significance and then builds a hierarchical tree of the identified structures. Here, the dendrogram {\it leaves} correspond to the cores in our analysis. We apply the dendrogram algorithm to the  3-$d$ H$_2$ number density, $n_{\rm H_2}$, which is derived using the local neutral gas fraction and  molecular gas mass calculated by {\small GIZMO}. Therefore, we ensure that identified cores are not only relatively dense but contain a significant amount of molecular gas. While these cores are not directly comparable to observed cores, this core definition returns structures that are reasonably likely to be detected in dense molecular gas tracers like NH$_3$ \citep[e.g.,][]{ChenPineda2019a,KerrKirk2019a}. 

The {\it dendrogram} tree is constructed using a nearest-neighbor search, which is agnostic about the underlying data structure, as well as about physics and morphology. This allows us use the native simulation resolution rather than ``flattening" the data to a uniform spatial grid.
This approach effectively isolates the structures of interest, cores, without interpolation or loss of resolution. We adopt a peak threshold, $min\_value = n_{\rm H_2} = 10^4$\,cm\eee, require structures to have a difference of at least $ 10^4$\,cm\eee\ in density between their peak and the density of parent node ($min\_delta$) and require that structures contain at least 100 cells ($min\_npix$). We mask out any cells with masses $\Delta m <10^{-3}\msun$, which represent recently ejected stellar feedback material. Thus, the identified cores effectively have a minimum mass of 0.1\,$\msun$.
We identify cores in each of the available snapshots (typically $>$500 snapshots per run), which have a uniform spacing of $\sim 24$ kyr.


We classify cores as protostellar if there is a protostar within 0.04 pc of the location of the peak density. We adopt a spatial criterion rather than a boundedness criterion since the latter is sensitive to the location of the core boundary and is influenced by the multiplicity of the protostellar system (tight binaries may have relatively high velocities). While some cores may be misclassified as protostellar under this criterion due to passing, unassociated interlopers we find good agreement between the classified protostellar cores and various discriminating host core properties, such as peak density (see \S\ref{sec:coreclusters}).

In this work we are particularly interested in the subset of cores that are long-lived and star-forming. To identify these, we follow the core gas over time and make an additional cut, as described in \S\ref{sec:tracking}, using the core evolutionary tracks to select for the cores of interest.

\subsubsection{Core Tracking}\label{sec:tracking}

We use the cell IDs to match and track structures over time. For each consecutive pair of snapshots we compare the gas cells assigned to each dendrogram structure.  
Given core $C$ in a snapshot at time $t$ and a core $C'$ in the following snapshot at $t+\Delta t$, we match the two cores if $N(C \cap C')/ N(C) > 0.5$ or if $ N(C \cap C')/ N(C') > 0.5$, where $N$ represents the number of cells. Note that core $C$ may be matched to more than one core at $t+\Delta t$, which indicates a {\it split}; similarly more than one core at time $t$ may be matched to $C'$, representing a {\it merger}. We do not require mergers and splits to be 1:2, such that a given core may split or merge into arbitrarily many cores. Although in practice splits or mergers involving more than 2 parents or children are very rare. 

This approach returns a set of core histories, where each track includes a single core in any given snapshot and each merger or split is recorded as a unique path. Consequently, tracks that merge or split will include the same cores for the overlapping (pre-split or post-merger) portion of their evolution. This approach is more sophisticated than that of \citet{OffnerTaylor2022a}, which did not take into account splits and used the center-of-mass in the tracking. Our approach is also more exact than the methods of either \citet{OffnerTaylor2022a} or \citet{SmullenKratter2020a} since we use cell IDs, rather than cell positions, to build the core history.

After the tracks are constructed,  we make several cuts and draw distinctions between different types of paths. First, we identify cores that are not connected to any tracks, i.e., the single-snapshot cores that have a lifetime $t < 2\Delta t \simeq 48$ kyr.  We remove these transient cores from the later analysis.  Next we identify the subset of tracks that have well-behaved star-formation histories i.e.,  ``good paths." These paths are either starless for their entire history or become protostellar monotonically. Such well-behaved protostellar tracks are initially starless, become protostellar once and remain protostellar until the core disperses (the track ends).  We exclude tracks that are not well-behaved and that bounce between protostellar and starless status from the analysis in \S\ref{sec:results}.
For the fiducial run, this process returns 79,167  well-behaved tracks, 6,021 (7.6\%) of which are star-forming. Of the protostellar tracks, there are 1,661 unique terminal protostellar cores, which is consistent with the number of stellar systems in the simulation. This indicates that each protostellar track splits or merges on average  $(6021 - 1661)/ 1661 \sim$ 2.6 times,  as every non-unique terminal core indicates a split or merger in the earlier history. We present a detailed discussion of the paths and their properties in \S\ref{sec:paths}.

\begin{figure}
    \centering
    \includegraphics[width = .47\textwidth, keepaspectratio]{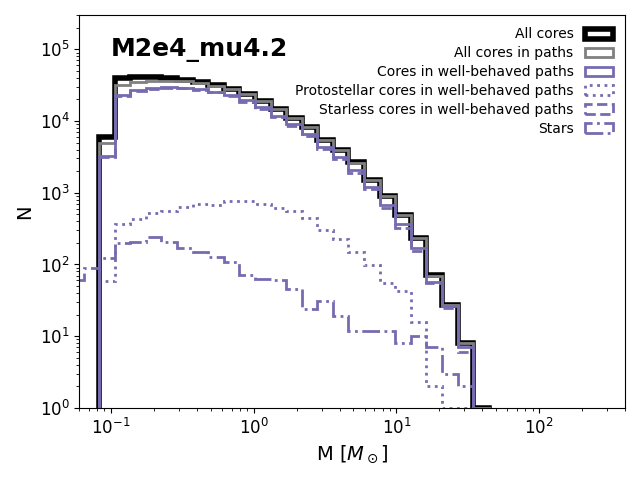}
    \includegraphics[width = .47\textwidth, keepaspectratio]{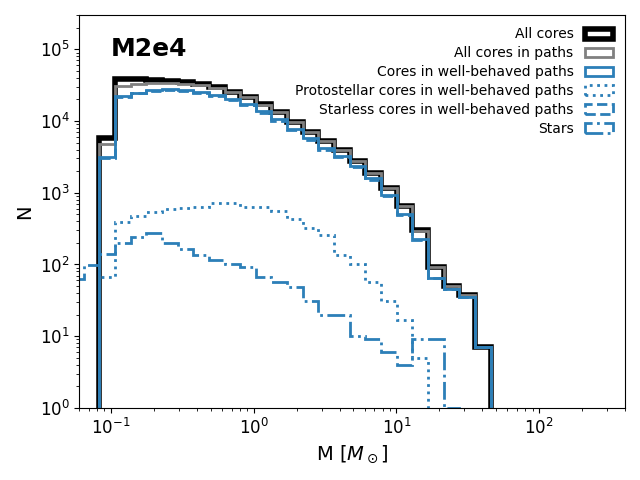}
     \includegraphics[width = .47\textwidth, keepaspectratio]{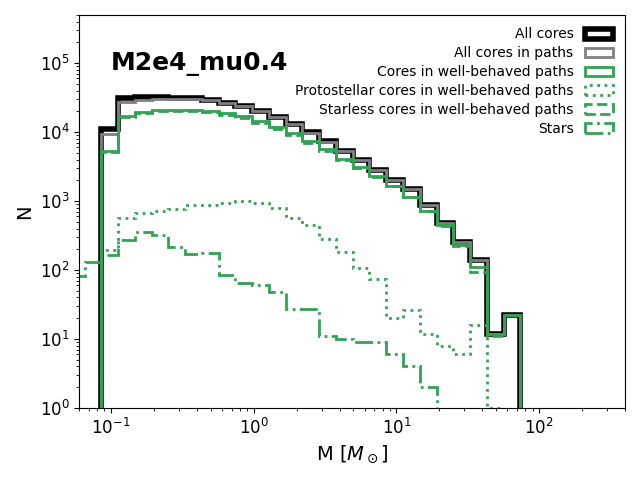}
 \caption{Distribution of core masses for simulation \lo (top), \med (middle), and \hi (bottom).  Lines indicate all identified cores (black), cores belonging to any path (gray), cores found in well-behaved  paths (solid color), protostellar cores in well-behaved paths (dotted color), starless cores in well-behaved paths (dashed color), and stellar masses (dash-dot color).
 For reference, \med\ has 363,331 identified cores, 339,114 cores in paths, 269,078 cores in well-behaved paths, 7,919 well-behaved protostellar cores, 261,159 well-behaved starless cores  and 2,246 stars. See Tables \ref{tab:IC_phys} and \ref{tab:cores} for core summary statistics.}
    \label{fig:cmf}
    \end{figure}

Figure \ref{fig:cmf} shows the distribution of core masses for all cores, only those in well-behaved paths, and protostellar cores. The single-snapshot cores and cores in paths that are not well-behaved are systematically lower mass. After these cores are removed, the core mass function is approximately log-normal with a peak around 0.4\,$M_\odot$. The full distribution of cores reaches a maximum around 0.3\,$M_\odot$.  If the dendrogram leaves reflect the underlying distribution of turbulent fragmentation, then the mass distributions should resemble a log-normal 
\citep{PadoanNordlund1997a,Hopkins_2012,GuszejnovGrudic2021a}. 


By comparison, the protostellar core distribution is flatter and has a peak around 0.6 $M_\odot$, an unsurprising shift given that protostellar cores are more likely to be gravitationally bound and hence more massive.  The median core masses of all distributions are comparable to or smaller than the Jeans mass ($M_J \sim 1 \msun$ for 10$^4$ cm\eee); the majority of the cores are not self-gravitating and thus the dendrogram is not preferentially selecting gravitationally fragmenting structures, which enables us to have a clearer and less-biased view of the dense gas evolution. The core masses are all systematically shifted compared to the stellar masses, likely due to a combination of fragmentation and stellar feedback entraining and expelling dense gas \citep[e.g.,][]{OffnerChaban2017a,GuszejnovGrudic2021a}.

\subsubsection{Constructing Core Properties}\label{sec:construct_core}


We compute a set of fundamental characteristics for each identified structure and use these to build a comprehensive view of core evolution. We represent each core by a vector of $d=25$ properties. These include fundamental bulk properties such as core mass, $M_c$, bulk velocity, $v_b$ (center-of-mass velocity), velocity dispersion, $\sigma_{\rm rms} = ( \Sigma_i m_i ( \mathbf{v}_i - \mathbf{v}_b )^2/M_c)^{0.5}$, 
effective radius $R_{\rm eff} = 5/3 \Sigma_i (m_i r_i^2)/M_c$, half-mass radius $R_h$, mean gas sound speed $c_s$, peak gas density, $n_{\rm max}$, and mean magnetic field, $\bar B$.  
 We also include the total kinetic energy, $E_{\rm K} = 0.5 \sum_i m_i (\mathbf{v}_i - \mathbf{v}_b)^2$, magnetic energy $E_B = (\Sigma_i m_i/\rho_i \mathbf{B}_i^2) / (8 \pi)$, gravitational potential energy, $E_{\rm grav}$, combined kinetic and thermal energy, virial ratio $\alphaturb = E_{\rm K}/ E_{\rm grav}$, and magnetic virial ratio $\alpha_{\rm B} = E_{\rm B}/ E_{\rm grav}$. 
 We use the {\it pytreegrav} package \citep{Grudic_2021pytree}  to compute the gravitational potential energy of each core, $E_{\rm grav}$, which does not {\it a priori} assume any specific geometry.

We include two parameters that describe the core shape. We carry out a principal component analysis of each leaf to determine the principal axes and then compute the major and minor aspect ratio, $R_b/R_a$ and $R_c/R_a$. Cores with small aspect ratios ($<$ 0.3) are essentially filaments. We also include the median outflow fraction as a measure of the influence of stellar feedback. The outflow fraction is the fraction of outflow material (by mass) in each cell.

Following \citet{OffnerTaylor2022a} we construct the density and velocity dispersion profiles using a set of 20 logarithmically distributed density bins over the range $n_{\rm H_2}=10^{3.5}-10^{7}$ cm\eee. 
We use the velocity dispersion profile to determine the radius of coherence, $R_{\rm coh}$, where $\sigma_{\rm rms}(R_{\rm coh})=c_s = 0.2$\,km s\e. This is similar to the observational definition for coherent cores \citep[e.g.,][]{ChenPineda2019a}, where this sound speed is comparable to the typical simulation core sound speeds (see \S \ref{sec:results}). In lieu of the density profile measurements we adopt the peak density and the slope of the density,  $\rho \propto r^{-p}$, as determined from a least-squares fit to all points in the density profile that fall between $r=0.0025$ pc and $r=0.1$ pc. Since cores may be either dense and compact or extended and fluffy, some fits may only span a few points at either end of this size scale. 
We use principal components analysis to down-select the velocity dispersion and density profiles to such that 95\% of its variance is retained in the PCA scores. This process selects four density and three velocity dispersion profile points, producing a total property vector of size 25 for each core. 


\subsection{Clustering Methodologies}\label{sec:cluster}

Following \cite{merenyi2017som} we invoke a hybrid prototype learning + graph-based clustering strategy. This is done partly to alleviate computational bottlenecks with the sheer number of cores to be clustered ($N > 2\times 10^5$ per run), but moreso to take advantage of the increased signal-to-noise resulting from prototype-based learning and to harness the ability of graph-based clustering to identify complex structures in data that (may) lie on sub-dimensional manifolds \citep{vathy2013graph}. 

\subsubsection{Prototype-Based Machine Learning}

Prototype-based machine learning models \citep{biehl2016prototype} learn intelligently formed representatives of the data called \textit{prototypes} that are then used downstream in subsequent machine learning tasks (e.g., classification, clustering). From a sample $X = \{x_i \in \mathbb{R}^d \}_{i=1}^N$ of size $N$, $M << N$ prototypes $W = \{ w_j \in \mathbb{R}^d \}_{j=1}^M$ are formed to best represent the data based on some goodness of fit criteria. Prototypes are generally found such that they (approximately) minimize \textit{quantization error}, or the error made when each datum is \textit{quantized} (represented by) its closest matching prototype. Common algorithms for obtaining prototypes include K-means \citep{macqueen1967} and more robust methods such as the Self-Organizing Map \citep{kohonen2001} or Neural Gas \citep{MartinetzSchulten1991}. In this work we formed prototypes of core properties via K-means. The number of prototypes chosen to represent $X$ impacts the amount of information in $X$ that can be represented by $W$.  Shannon's Rate-Distortion theory \citep{cover1991information} guarantees quantization error decreases monotonically as $M$ increases and can provide a lower bound on $M$ for a given maximum acceptable  quantization error, such a tolerance is rarely known a-priori. Instead, we invoke a rule of thumb \citep{arbelaitz2013extensive} and set $M = \mathcal{O}(\sqrt{N})$; specifically, to allow influence from our data's dimensionality $d$ (and not just sample size $N$), we set $M = N^{1/2} \times d^{1/4}$. In addition to sample size reduction ($N \to M$) prototype learning also boosts signal-to-noise ratios in $W$, especially as data dimensionality increases, via the benefits from the so-called Vector Quantizer Advantage \citep{lookabaugh1989high}.

\subsubsection{Graph-Based Clustering with Prototype Methods}

Aside from sample size and noise reduction, vector quantization provides a unique prototype similarity measure known as the CONNectivity Graph (CONN, \cite{tasdemir2009exploiting}) which posits prototypes as graph vertices and a graded similarity representing the local distribution of manifold neighbors as edge weights. Formally, $CONN_{ij} = \sum\limits_{i=1}^N I\left[ BMU1(x_i)=i \land BMU2(x_i)=j\right] + \left[ BMU1(x_i)=j \land BMU2(x_i) = i\right]$ where $I()$ is the indicator function and $BMU1,BMU2$ are the first and second Best Matching Units (closest prototypes in $W$) to datum $x_i$. Often, graph-based clustering methods invoke some type smoothed $k-$nearest neighbor graph to transition the view of data from Euclidean space to that of graphs. Such graphs represent purely \textit{distance} based, whereas CONN is an (un-normalized, local) \textit{density} based similarity. Benefits of using CONN over distance based similiarities in a variety of graph community detection (i.e., clustering) algorithms \citep{fortunato2010community} are shown in \cite{merenyi2017som}. Following the method of \cite{merenyi2017som}, once K-means prototypes of our data are learned they are clustered via the Walktrap algorithm \citep{pons2006computing} with default parameters (number of Markov Chain steps = 4). Walktrap is related to Spectral Clustering \citep{von2007tutorial} in that they both seek to minimize the Normalized Cut (Ncut) criteria of a graph \citep{shi2000normalized} (essentially, grouping nodes of high similarity into the same cluster, while minimizing cross-cluster similarities), but Walktrap approximates a minimal Ncut via graph Modularity \citep{fortunato2010community} maximization, whereas Spectral methods rely on analysis of the eigendecomposition of the graph's Laplacian, which can be imprecise for large and/or noisy data \citep{von2007tutorial}.

\section{Results}\label{sec:results}

\begin{table*}
\footnotesize
\hspace{-2cm}
	\begin{tabular}{|c|lccccccc|cc|}
	     \multicolumn{11}{c}{}\\ 
		 \multicolumn{1}{c}{}&
		 \multicolumn{8}{c}{\bf Properties of Cores in Well-Behaved Paths} &
		 \multicolumn{2}{c}{\bf Fits}\\
		\hline
		{\bf Cloud} & Type & $N_c$ & $M_c$($\msun$) & $R_c$(pc) & $\sigma_{\rm rms}$(km/s) & $c_s$(km/s) &
  $\alphaturb$ & $\alpha_B$ & $M_c \propto R_c^{p_m}$ & $\sigma_{\rm rms} \propto R_c^{p_v}$    \\
		\hline 
  \lo & All & 290,859 & 0.39$_{-0.18}^{+0.42}$ & 0.08$_{-0.03}^{+0.06}$ & 0.36$_{-0.12}^{+0.19}$ & 0.21$_{-0.05}^{+0.05}$ & 3.13$_{-1.87}^{+4.93}$ & 3.02$_{-1.76}^{+4.29}$ & 1.052$\pm$0.002 & 0.254$\pm$0.002  \\ 
   & Starless & 282,190  & 0.39$_{-0.18}^{+0.41}$ & 0.08$_{-0.03}^{+0.06}$ & 0.36$_{-0.12}^{+0.19}$ & 0.21$_{-0.05}^{+0.05}$ & 3.23$_{-1.91}^{+5.04}$ &  3.16$_{-1.81}^{+4.36}$ & 1.064$\pm$0.002 & 0.269$\pm$0.002 \\
    & Protostellar & 8,669 &  0.71$_{-0.39}^{+0.79}$ & 0.04$_{-0.02}^{+0.03}$ & 0.42$_{-0.12}^{+0.18}$ & 0.20$_{-0.04}^{+0.05}$ & 1.15$_{-0.52}^{+1.29}$ & 0.37$_{-0.14}^{+0.29}$ & 0.689$\pm$0.007 & -0.051$\pm$0.007 \\ 
    \hline 
    \med & All & 269,078 & 0.39$_{-0.18}^{+0.43}$ & 0.08$_{-0.03}^{+0.06}$ & 0.33$_{+0.11}^{+0.17}$ & 0.20$_{-0.04}^{+0.05}$  & 2.43$_{-1.40}^{+3.77}$ & 2.49$_{-1.39}^{+3.35}$ &  1.099$\pm$0.002 & 0.239$\pm$0.002 \\ 
(fiducial)   & Starless & 261,159 & 0.38$_{-0.18}^{+0.42}$ & 0.08$_{-0.03}^{+0.06}$ & 0.33$_{-0.11}^{+0.16}$ & 0.20$_{-0.04}^{+0.05}$  & 2.49$_{-1.44}^{+3.84}$ & 2.60$_{-1.42}^{+3.41}$ &  1.109$\pm$0.002 & 0.257$\pm$0.002  \\
    & Protostellar & 7,919 &  0.61$_{-0.33}^{+0.69}$ & 0.03$_{-0.01}^{+0.03}$ & 0.42$_{+0.12}^{+0.19}$ & 0.21$_{-0.04}^{+0.04}$  & 1.14$_{-0.53}^{+1.28}$ & 0.36$_{-0.13}^{+0.32}$  &  0.752$\pm$0.007 & -0.045$\pm$0.007 \\ 
    \hline
    \hi & All & 224,621 & 0.48$_{-0.24}^{+0.63}$ & 0.09$_{-0.03}^{+0.08}$ & 0.28$_{-0.09}^{+0.15}$ & 0.19$_{-0.04}^{+0.04}$ & 1.51$_{-0.81}^{+2.23}$ & 1.91$_{-0.99}^{+2.14}$ & 1.207$\pm$0.002 & 0.335$\pm$0.002 \\ 
   & Starless &214,526  & 0.47$_{-0.24}^{+0.62}$ & 0.09$_{-0.04}^{+0.08}$ & 0.27$_{-0.08}^{+0.15}$ & 0.19$_{-0.04}^{+0.04}$ & 1.54$_{-0.83}^{+2.28}$ & 2.01$_{-1.01}^{+2.21}$ & 1.208$\pm$0.002 & 0.359$\pm$0.002 \\
    & Protostellar & 10,095 &  0.63$_{-0.35}^{+0.66}$ & 0.04$_{-0.02}^{+0.03}$ & 0.38$_{-0.10}^{+0.15}$ &  0.19$_{-0.03}^{+0.04}$ & 1.09$_{-0.51}^{+1.20}$ & 0.41$_{-0.16}^{+0.39}$ & 1.016$\pm$0.005 & -0.020$\pm$0.007  \\
    \hline    

\end{tabular}
        \vspace{-0.1cm}
 \caption{Summary statistics for all cores in well-behaved paths. The first set of columns lists the core median properties by class: mass, radius, velocity dispersion,  sound speed, 
 turbulent viral parameter, magnetic virial parameter. The ranges indicate 25\% quartiles. The second set of columns lists the least-squares fits of mass versus radius and velocity dispersion versus radius for the different core populations.} 
 \label{tab:cores}
\end{table*}

\begin{table}
\hspace{-2cm}
	\begin{tabular}{|c|ccc|c|}
	     \multicolumn{4}{c}{}\\ 
		 \multicolumn{1}{c}{}&
		 \multicolumn{3}{c}{\bf Path Statistics} &  \multicolumn{1}{c}{}\\
		\hline
		{\bf Cloud} & Type & $N_{\rm path}$ & $ t_{\rm life}$ (Myr) & $t_{\rm proto}$ \\
		\hline 
  \lo & All & 78,469 & 0.49$_{-0.35}^{+0.74}$  &\\ 
   & Starless  & 73,210 &  0.47$_{-0.32}^{+0.72}$ &  0.10$_{-0.05}^{+0.15}$ \\
    & Protostellar & 5,259 & 0.92$_{-0.59}^{+1.01}$ & \\
    \hline 
    \med & All & 79,167 & 0.64$_{-0.47}^{+0.94}$  &\\ 
  (Fiducial) & Starless & 73,146 & 0.62$_{-0.45}^{+0.92}$ & 0.10$_{-0.05}^{+0.12}$ \\
    & Protostellar & 6,021 & 1.09$_{-0.74}^{+1.14}$ & \\ 
    \hline
    \hi & All & 45,293 & 0.49$_{-0.35}^{+0.87}$ & \\ 
   & Starless & 41,024 &  0.47$_{-0.35}^{+0.84}$ &  0.12$_{-0.07}^{+0.12}$\\
    & Protostellar & 4269 & 0.79$_{-0.45}^{+0.94}$ &  \\
    \hline 
    \hline
\end{tabular}
        \vspace{-0.1cm}
 \caption{Summary statistics for  well-behaved paths. The columns are simulation name, type of path, number of paths, median path lifetime with 25\% quartile ranges, and median length of the protostellar phase.  }
 \label{tab:paths}
\end{table}

\subsection{Core Properties}


The number of identified cores varies significantly with time and simulation properties. At early times, while over-densities are still growing from the initial turbulent perturbations, there are few structures and those that are identified tend to be large. These gradually break into smaller substructures, a small subset of which eventually gravitationally collapse. The number of cores declines at late times as feedback disperses the cloud. 

As summarized in Table \ref{tab:IC_phys} across all snapshots the runs have 391,261 (\lo),  363,331 (\med), and  333,108 (\hi) total cores with the number declining by 15\% with increasing magnetic field strength. We note that the core number decreases less steeply than the overall star formation rate, which declines by $\sim$40\% (from 9\% to 5\%) with increasing field strength (see Table \ref{tab:IC_phys}). This indicates that the magnetic field, while suppressing collapse, does not significantly reduce the number of cores. However, in the \hi\ run, the cores that are identified are more likely to be protostellar, 13\% versus $\sim7$\% in the other two runs, in part because this runs has a higher number of protostars.

\subsubsection{Core Properties as a Function of Path Membership}


 In all cases the large majority of identified cores disperse before forming stars. However, a small subset are quite short lived: about 4-7\% of cores in each run survive less than $2\Delta t$ and are not connected to any path. 
 The top panels in Figure \ref{fig:coreprop} show that these cores are less massive ($M \lesssim 1 \msun$) and have systematically higher velocity dispersions and virial ratios. These characteristics are consistent with being less gravitationally bound than their longer-lived counterparts. However, all cores, independent of path membership, exhibit approximate energy equipartition such that the kinetic and magnetic ratios are comparable, a trend that spans three orders of magnitude in $\alpha_{\rm vir}$ and $\alphaB$. Altogether, there is significant overlap between the transient, pathless cores and those belonging to paths, which indicates that the core properties are correlated with but not predictive of core outcomes.

 Approximately 19-29\% cores in each run belong to complex paths, which we remove from the sample.
These cores are less distinct from the cores found in well-behaved paths. This is perhaps unsurprising as many of these cores are participating in star-formation activity and hail from messy protostellar paths that have a significant number of splits and mergers. Figure \ref{fig:coreprop} shows that cores in complex paths have similar masses, sizes, and velocity dispersions to cores in well-behaved paths, although they have a slight trend to lower virial ratios. Unlike the pathless cores, the populations of cores in paths exhibit a correlation between mass and size ($M \propto R^{0.6-1.2}$) and linewidth and size ($\sigma \propto R^{0.24-0.36}$). Table \ref{tab:cores} presents the fits for the well-behaved core populations. 
 
 We find that the core distributions and property relationships are relatively similar across all three runs; Figures \ref{fig:coreprop_lo} and \ref{fig:coreprop_hi} in the appendix show the core distributions for runs \lo\ and \hi.

\subsubsection{Starless and Protostellar Cores in Well-behaved Paths}

 Table \ref{tab:cores} summarizes the properties of cores in well-behaved paths. Protostellar cores comprise only $\sim$3-4.5\% of the population of well-behaved cores. However, there are fewer protostellar cores in well-behaved paths in the weaker magnetic field runs, suggesting that the length of the protostellar phase is shorter (see \S\ref{sec:paths}). Meanwhile, the total number of starless cores declines with increasing magnetic field strength, since higher magnetic pressure inhibits more structures from reaching the 10$^4$cm$^{-3}$ identification threshold. The difference is more dramatic when considering that the overall cloud lifetime increases with magnetic field strength such that the cloud disruption time in almost 40\% later in run \hi~(see Table \ref{tab:IC_phys}).

\begin{figure*}
    \centering
    \includegraphics[width = 1.0\textwidth, keepaspectratio]{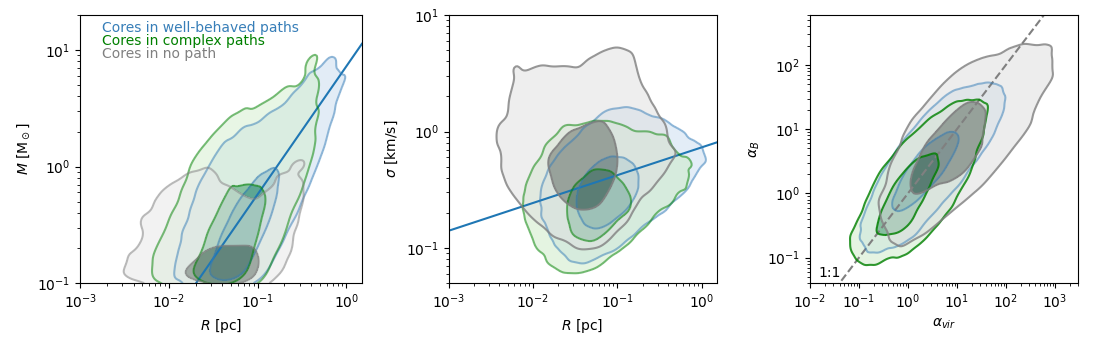}
    \includegraphics[width = .95\textwidth, keepaspectratio]{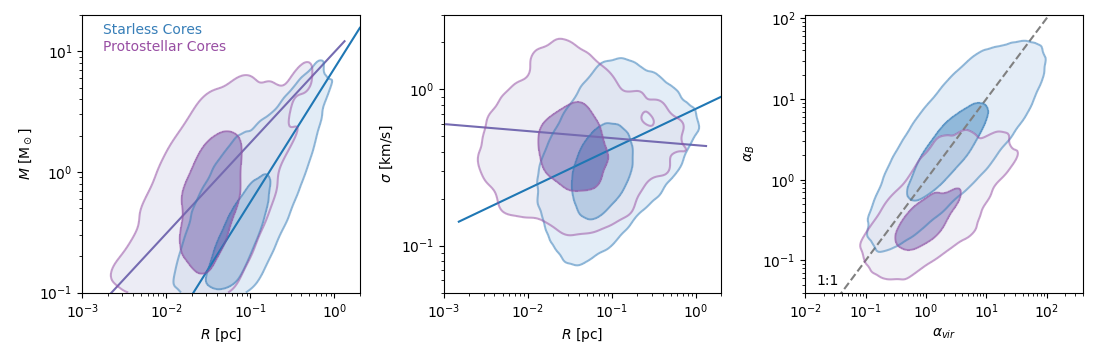}
 \caption{Distribution of core properties for simulation \med. From left to right: core mass versus effective radius, velocity dispersion  versus effective radius, and magnetic virial parameter versus turbulent virial parameter. Top row: Cores in well-behaved paths are blue, cores in paths that behave non-monotonically are in green, and identified cores that are unconnected to any path are in gray. Bottom row: starless cores in well-behaved paths are in blue and protostellar cores in well-behaved paths are purple.  The outer contours enclose 95\% of the data; the inner contours encloses 50\% of the data.} 
    \label{fig:coreprop}
    \end{figure*}

The populations of starless and protostellar cores display significantly different mean properties. The bottom panels in Figure \ref{fig:coreprop} show that starless cores are systematically larger and have higher virial ratios than protostellar cores. Protostellar cores have a slightly steeper mass-size relationship ($M \propto R^{0.7}$) but show no correlation between velocity dispersion and size. This is likely because the velocity dispersions of many protostellar cores are dominated by motions driven by stellar feedback and collapse rather than cascading cloud turbulence. Like starless cores, protostellar cores follow the 1-1 line of energy equipartition but have lower virial ratios and are offset to lower magnetic virial ratios, a trend consistent with lower magnetic pressure support and gravitational instability.

While there is significant overlap between the starless and protostellar core property distributions, Table \ref{tab:cores} indicates that a typical protostellar core is nearly twice as massive and half the size of a typical starless core; protostellar cores are much more compact and have higher mean densities than starless cores. While the median velocity dispersion of protostellar cores is only $\sim$20\% higher, the median kinetic virial ratio is up to 3 times smaller and magnetic ratios are up to $\sim 8$ times smaller than those of the typical starless core. It is not the case that protostellar cores are strongly bound, as the median $\alphaturb$ is slightly above unity; instead, starless cores are often highly turbulent and significantly unbound. The median values of $\alphaB$ indicate that magnetic support plays a significant role in resisting collapse, even for cores in the weakest magnetic field run \lo\ where the median starless magnetic ratio $\alphaB \sim 3$. 

While there is some variation between the three runs, starless and protostellar core masses, sizes, dispersions, and sound speeds are relatively similar between the runs. The protostellar cores in all three runs also have comparable virial ratios, irrespective of the cloud magnetic field strength. Some of this uniformity is likely due to the amplification of the magnetic field at high densities (see Fig.~12 in \citealt{Guszejnov_2022_imf}). In contrast, the starless core virial parameters vary with the initial cloud magnetic field: starless cores identified in run \hi\ have virial ratios nearly half that of their counterparts in run \lo. These starless cores are also {\bf slightly} more massive and slightly larger, indicative of higher magnetic support, than those in the \med\ and \lo\ runs,  but the difference is well within one standard deviation.

\subsection{Core Evolution} 

Cores in all simulations experience significant variation in their properties over time. Core merging and splitting is frequent. Some of this behavior is due to the dendrogram algorithm itself, which is sensitive to small changes in the local peak density \citep{SmullenKratter2020a}. However, a significant amount of this is simply the nature of turbulent gas: cores identified using the gas density do not have well-defined boundaries. These over-dense structures grow and accrete, advect with the local flow, and encounter shocks, which may disperse them. Nonetheless, tracking cells individually allows us to reliably connect identified structures across time, independent of their evolving properties, and measure their behavior and longevity within the larger cloud environment.

\begin{figure}
    \centering
     \includegraphics[width = .47\textwidth, keepaspectratio]{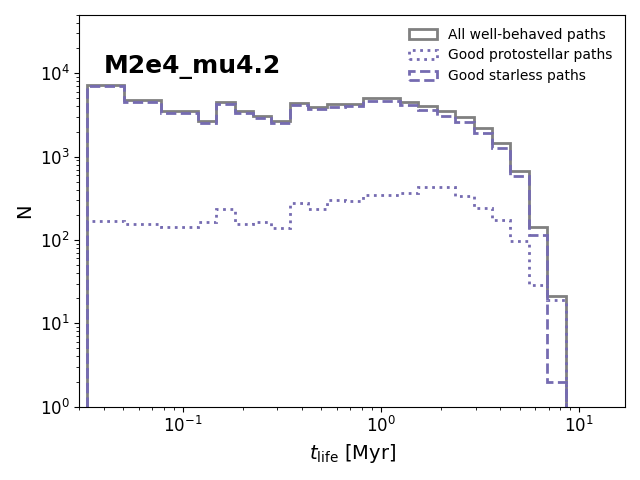}
    \includegraphics[width = .47\textwidth, keepaspectratio]{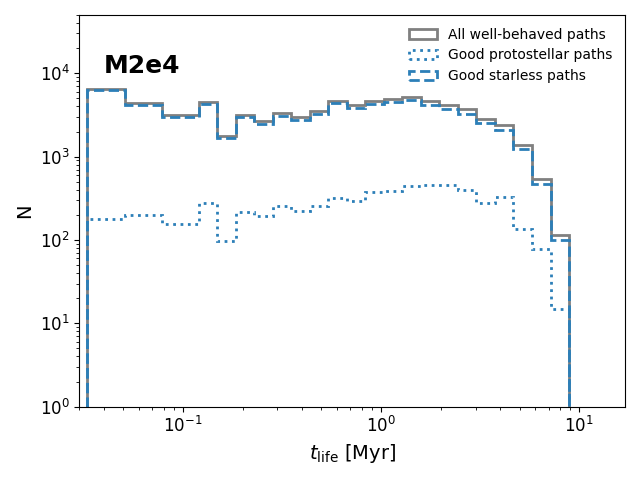}
     \includegraphics[width = .47\textwidth, keepaspectratio]{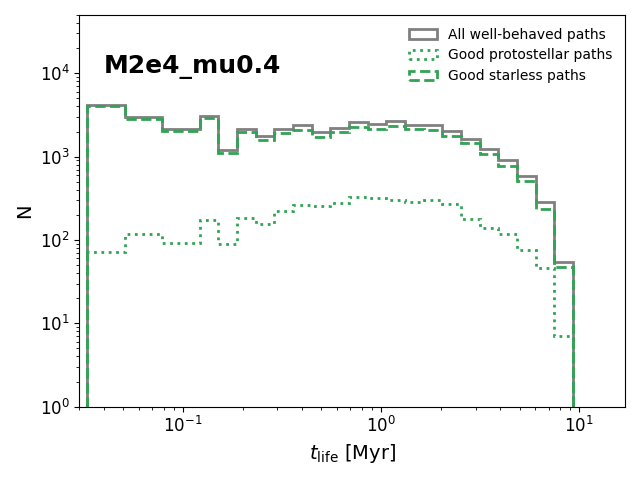}
 \caption{Distribution of path lifetimes for simulation \lo~(top), \med~(middle) and \hi~(bottom). The population of well-behaved paths is indicated by a solid grey line,  well-behaved protostellar paths are dotted, and well-behaved starless cores are dashed. }
    \label{fig:life}
    \end{figure}
    
\subsubsection{Statistics of Evolutionary Paths}\label{sec:paths} 

Table \ref{tab:paths} summaries the number of well-behaved paths and their median lifetimes. 
There are nearly 80,000 well-behaved paths in runs \lo\ and \med, while run \hi\ has $\sim$45,000 paths, despite the relatively longer simulation time. This difference in path number is partially due to the smaller number of structures but mostly due to the larger fraction {\bf of} protostellar cores, which are more likely to be connected by complex paths that we exclude.
In each run, only a few thousand ($\sim 6-8$\%) paths are protostellar.  

The median protostellar path lifetime is $0.8-1.0$ Myr, which is about 40\% longer than the median starless core lifetime.  There is no monotonic trend in the median lifetime with increasing magnetic field strength, generally due to the distribution being relatively flat in all cases as shown in Figure \ref{fig:life}. The distribution of starless paths appears slightly double peaked with a rising number of paths towards short lifetimes and a peak at $\sim 1$\,Myr. The distribution of protostellar paths exhibits a similar peak around $\sim$1\,Myr but no rise towards short lifetimes. The distributions of starless and protostellar core lifetimes are relatively similar, except for an excess of starless cores that are short-lived; these cores may partially explain the shorter mean lifetime of starless cores. The core lifetimes appear to be largely insensitive as to whether or not a star eventually forms.

The typical length of the protostellar path is several times longer than the core freefall time, 
where the core freefall time is $0.39^{+0.30}_{-0.18}$ Myr in the \med\ run. This is because the core lifetimes include the time to assemble the gas before collapse ensues, as well as the post-collapse evolution (see \S\ref{sec:evol} for more discussion).   We define the protostellar lifetime as the length of time during which the core is classified as protostellar, i.e., beginning at the snapshot when a protostar forms in the core. The protostellar paths spend on average $\sim 0.1$\,Myr in the protostellar phase, in all runs, independent of magnetic field strength (Table \ref{tab:paths}). This underscores the important and universal role feedback, outflows in particular, play in dispersing the core gas. Note that the end of the core lifetime marks the point at which either the protostar decouples from (leaves) the core or the mass of the natal gas structure falls below 0.1\,$\msun$. Accretion may continue if some residual gas lingers or the protostar enters another core.

Note that the number of protostellar paths is significantly higher than the number of stars formed in the calculations. Each of the runs has only $\sim$1,451-1,661 unique final protostellar cores, which is comparable to number of final star systems \citep{GuszejnovRaju2023a,FariasOffner2024a}. This indicates that the well-behaved protostellar paths experience at least three or four splits or mergers. While each split or merger event produces a new path, the well-behaved path list only includes those that split/merge starless-to-starless cores or protostellar-to-protostellar cores. Despite this stochastic behavior, the paths do exhibit some trends in their global properties, which we discuss in the next section.


\begin{figure*}
    \centering
    \includegraphics[width = .95\textwidth, keepaspectratio]{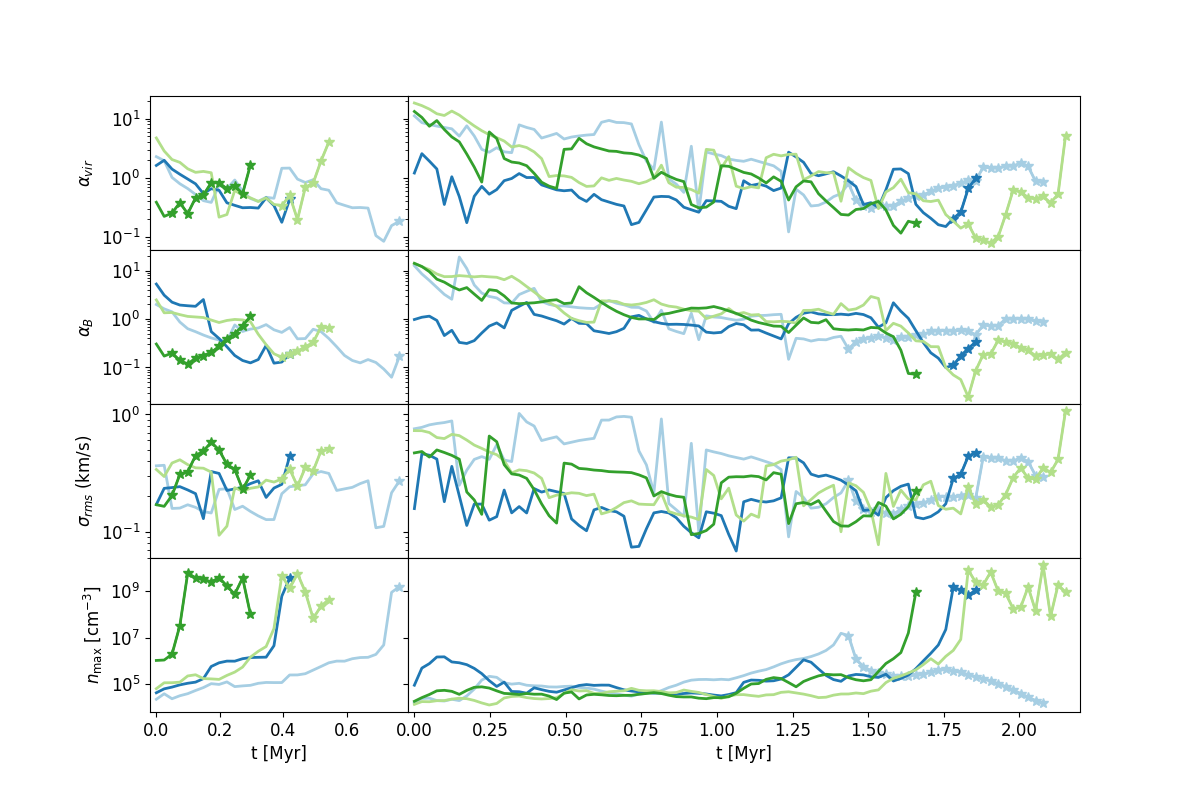}
 \caption{Time evolution of eight representative protostellar cores from \med: four short-lived cores (left) and four long-lived cores (right). The star symbols indicate the formation of a protostar. The panels from top to bottom display the viral parameter, magnetic virial parameter, velocity dispersion, and peak number density in the core.}
    \label{fig:coreevo}
    
    \end{figure*}

\begin{figure*}
    \centering
    \includegraphics[width = 1.0\textwidth, keepaspectratio, trim={1.9cm 0 1.9cm 0},clip]{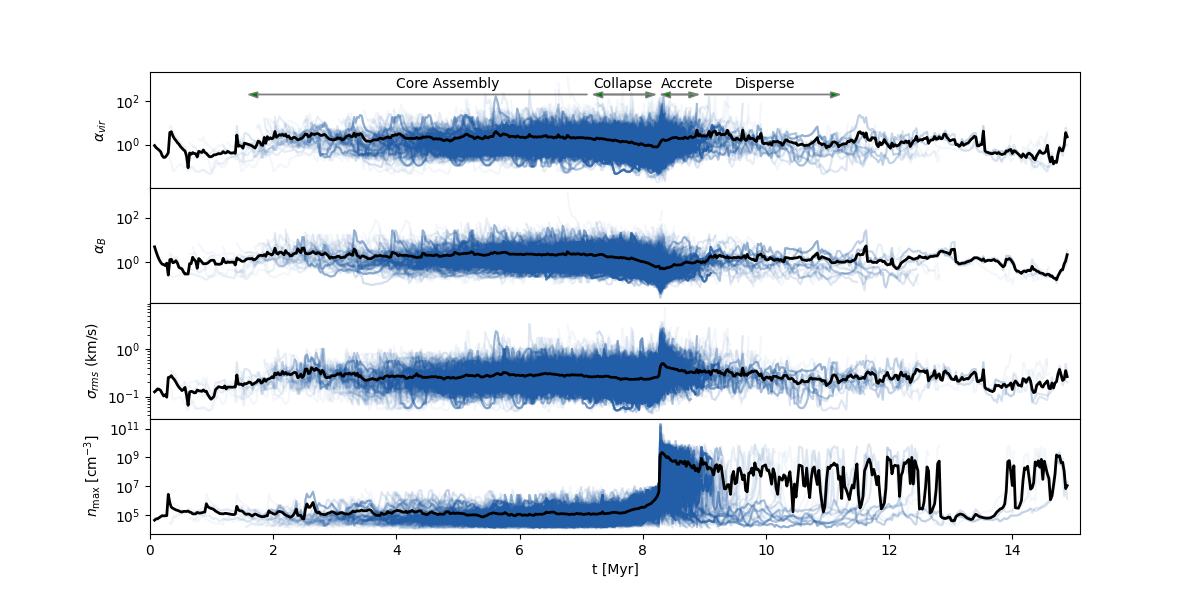}
 \caption{Time evolution of all protostellar cores from \med, where the solid black line indicates the mean across all tracks.  The panels from top to bottom display the viral parameter, magnetic virial parameter, velocity dispersion, and peak number density in the core. The tracks are matched to the time they reach $n=10^7$\,cm$^{-3}$; the time on the x-axis is shown for reference and does not correspond to the time the cores exist in the simulation. Four core phases are labeled in the top panel; the length of each phase varies between cores. } 
    \label{fig:allcoreevo}
    
    \end{figure*}

\subsubsection{Trends in Core Evolution}\label{sec:evol}

Figure \ref{fig:coreevo} shows the evolution of the virial ratio, velocity dispersion, and peak density for eight representative protostellar paths, four with lifetimes between 0.3-0.8\,Myr and four with lifetimes between 1.6-2.2\,Myr. The paths exhibit very heterogeneous evolution, and all parameters vary non-monotonically with time. Some of this variability is due to the dendrogram identification, but much of it is simply reflecting the nature of turbulent flow. Nonetheless, despite short timescale variation, the paths exhibit some global trends.  Figure \ref{fig:allcoreevo} shows these four properties averaged across all protostellar cores. We match the tracks to the point the core first reaches a density of $10^{7}$\,cm$^{-3}$, which highlights some systematic trends in the overall core evolution.  

The evolution can be considered to be composed of four components,  which are labeled on Figure \ref{fig:allcoreevo}: a core assembly stage, during which the included mass ebs and flows, a collapse stage, when gravitational instability occurs but before a protostar forms, an accretion phase during which the new protostar grows, and a dispersal phase, during which protostellar feedback destroys the remaining core. Not all protostellar paths clearly exhibit all of these stages, and each phase spans a broad range of timescales.

Figure \ref{fig:coreevo} shows that during assembly cores in the long paths experience a significant amount of stochastity in their properties as they collect gas, contract, and expand with passing shocks. The early evolution of these protostellar cores is similar to that of many cores that do not go on to form stars, i.e., the initial conditions are not deterministic \citep{OffnerTaylor2022a}.  The top panels of Figures \ref{fig:coreevo}  and \ref{fig:allcoreevo}  show that during this phase the virial parameters of these star-forming cores gradually decline as they lose magnetic support and become self-gravitating.
The turbulent velocity dispersion also declines until it becomes sonic, $\sigma_{rms} \lesssim c_s \sim 0.2$ cm s\e. 

Collapse ensues once  cores reach sufficiently low virial parameters.
The bottom panels in Figure \ref{fig:coreevo}  suggest that there is a critical central density beyond which collapse is irreversible. At this point, the density increases sharply over a freefall time of $\sim$0.1\,Myr.  While the critical density here appears to be around $n \simeq 10^6$\,cm$^{-3}$, we expect that the critical density leading to runaway collapse for a particular core depends on the local conditions and cloud properties \citep{PriestleyClark2023a, MoonOstriker2024a}.

After the protostar forms the cores may disperse immediately or persist for $\sim$0.5\,Myr. The behavior depends on the amount of protostellar core mass and the core environment. Cores with masses close to 0.1 Msun may quickly fall below the dendrogram identification threshold for cell number (e.g., dark green long path, light and dark blue short paths). Higher mass cores or cores that reside in a filamentary region can be sustained by accretion for quite some time (light blue long path). During the accretion phase, the kinetic virial ratio increases as outflows drive turbulence in the core. The magnetic virial ratio may also slowly rise, since magnetic flux is not accreted onto the sink particle along with the gas. Physically this treatment mirrors the relatively low magnetic energy of stars compared to their parent core gas.

Core dispersal happens relatively rapidly and often coincides with a sharp increase in velocity dispersion and kinetic virial parameter (e.g., light green and dark blue long paths). During this time the peak density may also decline as the core expands (e.g., light and dark green short paths). 

The dark green short path, which contains only a single starless core and appears to be protostellar before collapse begins, represents a piece from a larger, fragmenting protostellar structure.  
Similarly, the light blue long path also exhibits an evolutionary path that does not fully capture the collapse phase and likely represents a split from a larger protostellar structure. 
The complexity of these paths defies a simple narrative of protostar formation, e.g., as provided by the classic, isothermal sphere model \citep{Shu_1977}; here each evolutionary track has a unique story.

\begin{figure*}
    \centering
    \includegraphics[width = .7\textwidth, keepaspectratio]{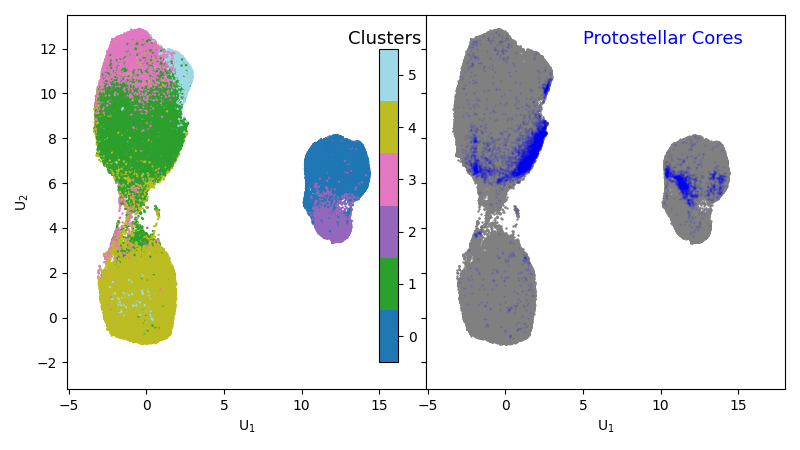}
    \includegraphics[width = .7\textwidth, keepaspectratio]{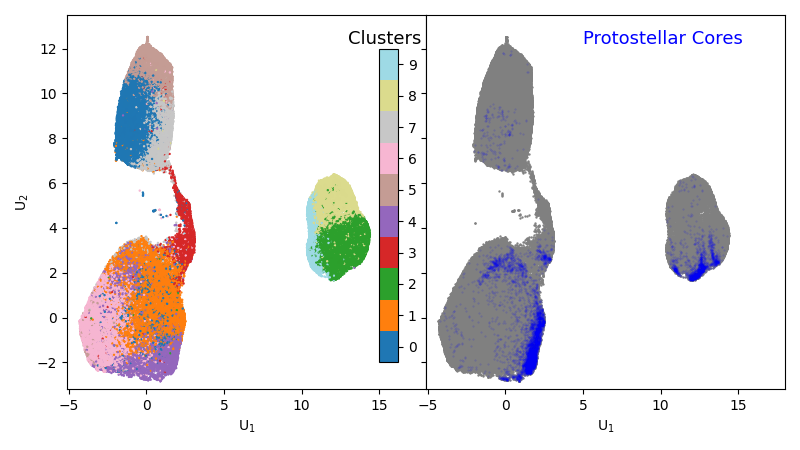}
    \includegraphics[width = .7\textwidth, keepaspectratio]{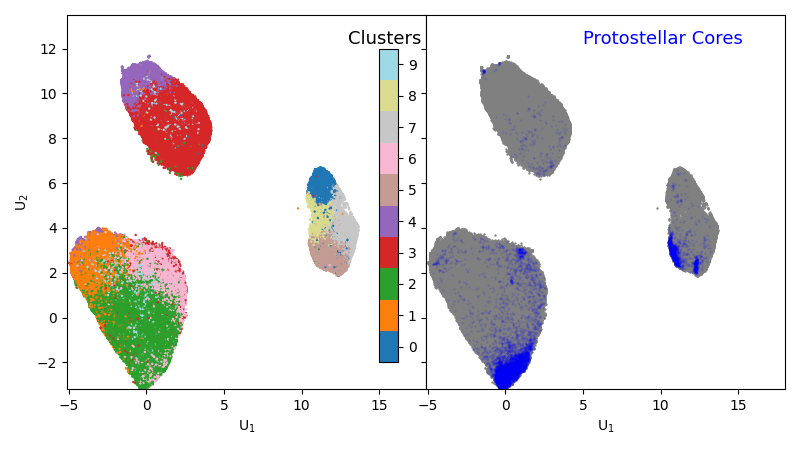}
 \caption{ UMAP projection of the \lo\ (top), \med\ (middle) and \hi\ (bottom) core properties, where each dot represents a core belonging to a well-behaved path. Left: colored by identified subgroups. Right: colored by core state, where starless cores are grey and protostellar cores are blue. The projected clusters are grouped into three main groups in the two-dimensional space. In the \med\ run the protostellar cores are predominantly members of clusters 1, 2 and 4.}
    \label{fig:umap_groups}
    \end{figure*}

\begin{figure*}
    \centering
    \includegraphics[width = .95\textwidth, keepaspectratio]{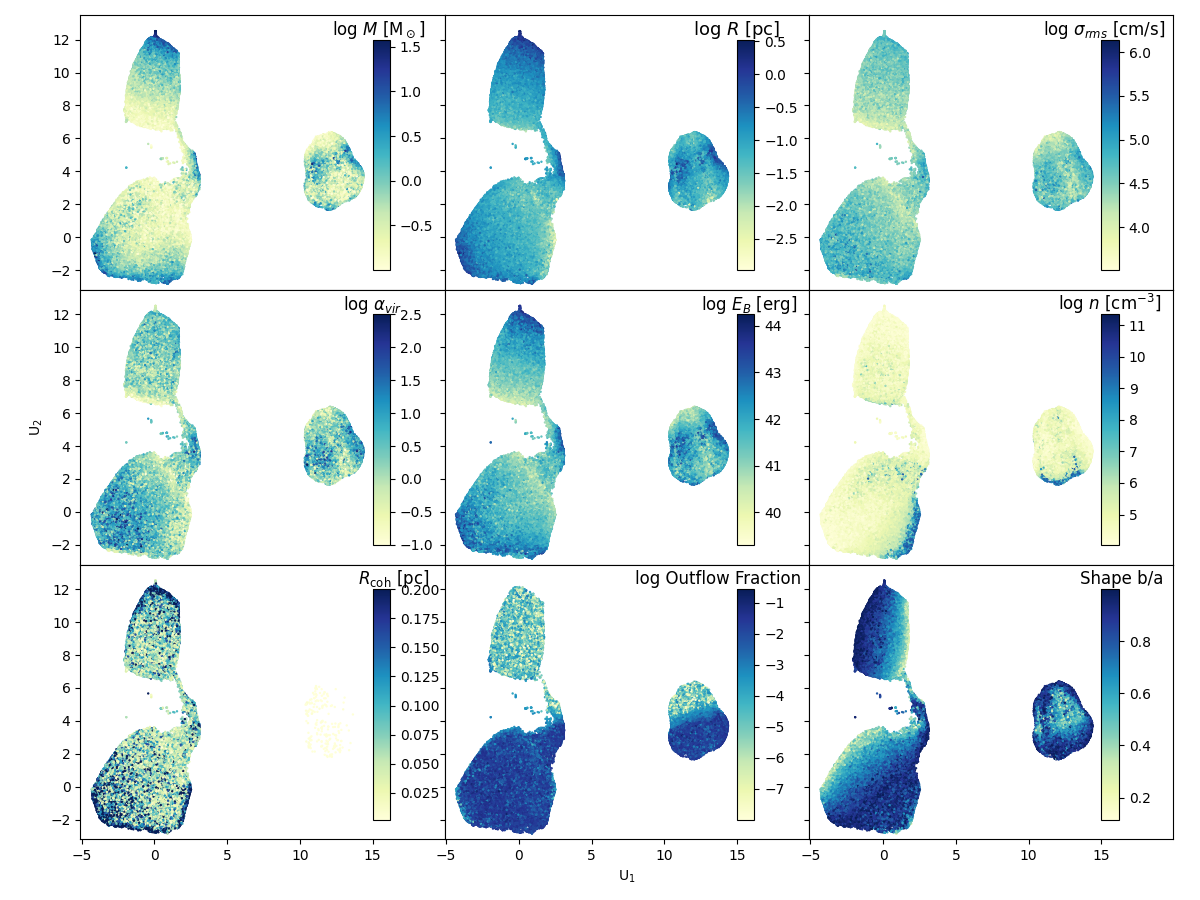}
 \caption{ UMAP projection of core properties from run \med, where each dot represents a core belonging to a well-behaved path. Top: UMAP colored by core mass (left), radius (center), and velocity dispersion (right). Middle: UMAP colored by virial parameter (left), magnetic energy (center) and peak density (right). Bottom: UMAP colored by the size of the region of coherence ($R_{\rm coh}$ defined as the outer radius where $\sigma_{\rm rms}(r) < c_s$, left), mean outflow mass fraction (center), and aspect ratio (right).  Cores that contain no coherent region at all are not included in the bottom left panel. The shape of the projected distribution is strongly influenced by whether cores are coherent (left structures) and whether they contain a significant fraction of outflow material (lower structures).  The bulk properties shape the ordering of the cores within each region.}
    \label{fig:umap_props}
    \end{figure*}

    \begin{figure*}
    \centering
    \includegraphics[width = .98\textwidth, keepaspectratio]{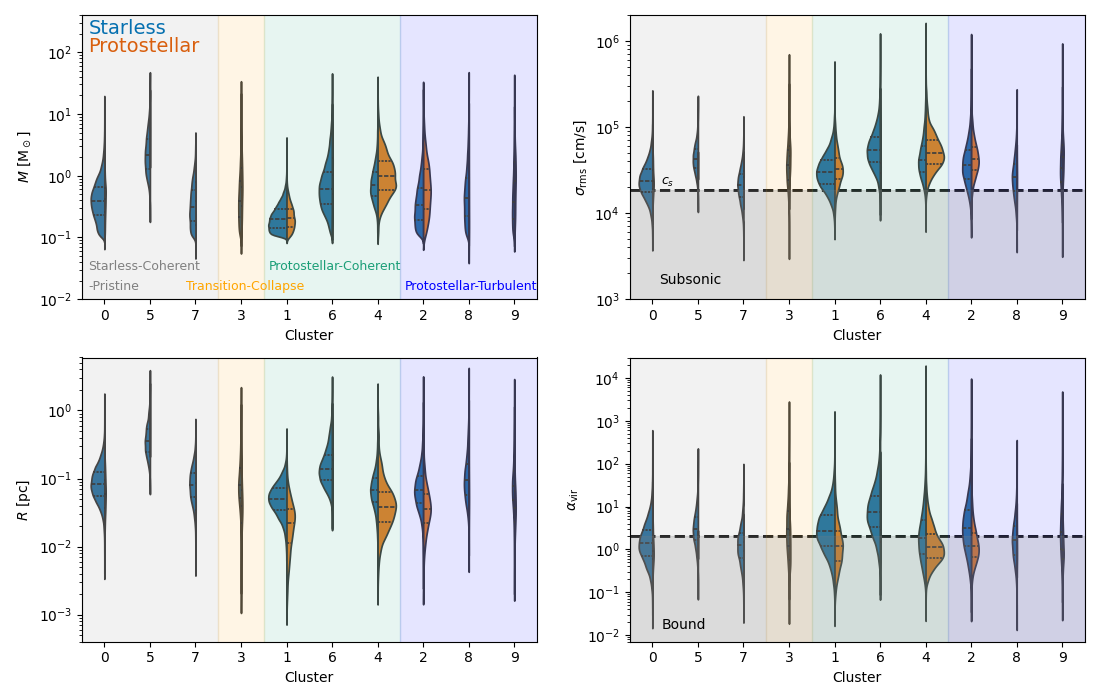}
 \caption{ Violin plots of the core properties for each cluster for run \med, where the distributions of all cores in well-behaved paths are shown in blue and the protostellar core distributions are orange. Shaded regions in each panel indicate the major UMAP groupings: top left ( predominantly  starless  cores where a significant number of cores have coherent centers), bridge (transitioning and/or collapsing cores), bottom left ( groups containing a significant number of  protostellar  cores and some cores with coherent centers), right (protostellar and/or turbulent cores that do not have coherent centers). Top left: core masses. Top right: velocity dispersion, where the sound speed of 10\,K gas is indicated by the dashed line; the shaded region below indicates cores with mean subsonic velocity dispersions. Bottom left: core radius. Bottom right: turbulent virial parameter, where the shaded region below the dashed line indicates gravitationally bound cores.}
    \label{fig:groups}
    \end{figure*}

\subsection{Core Clusters}\label{sec:coreclusters}

\subsubsection{UMAP Structure}

As described in \S\ref{sec:cluster}, we construct a representative vector of properties for each core, and then we cluster the vectors in high-dimensional space. Figure \ref{fig:umap_groups} displays the identified groups, where we use Uniform Manifold Approximation and Projection \citep[UMAP][]{mcinnes2018} to project and visualize the result in two-dimensions. We use a spectral embedding to initialize the low-dimensional space and adopt the default settings for the hyperparameters, e.g., {\it n\_neighbors} = 15, {\it min\_distance} =0.01, and {\it output\_metric = `euclidean'}.  Note that while the absolute UMAP coordinates are not meaningful, the relative spacing and positioning between the groups provides a measure of the distance between and similarity of the cores in the high dimensional space. 

The left panels of Figure \ref{fig:umap_groups} shows that the identified clusters map to distinct, if not always well-separated, portions of the UMAP space.
Likewise, the right panels of Figure \ref{fig:umap_groups} show that protostellar cores map to similar locations in the UMAP for all three runs. Since the protostellar flag is not contained in the vector used in the clustering, the separation is driven by other properties that correlate with protostellar cores, e.g., maximum density. 
Both of these factors give confidence that the clusters represent unique subsets of core properties.

All simulations exhibit three visually distinct groups of cores, i.e., a clustering applied in the two-dimensional UMAP space would likely return three clusters rather than the six or ten identified in the high-dimensional space. The UMAP projection indicates that each of these larger structures are composed of one to three subgroups identified from the high-$d$ property vector. The two visually distinct structures on the left in Fig.~\ref{fig:umap_groups} are closer together in run \lo; the same two structures are connected by a bridge that is identified as a distinct subgroup in run \med. In run \hi\ these two structures are completely separate, however, a small number of cores in the subgroups 2 (green), 3 (red) and 4 (purple)  cross over to the other structure, indicating that some connections between these structures remain in the high-dimensional space. 

All three runs also contain a third distinct structure on the right that is fully disconnected from the other two. No cores from any of the these subgroups appear in the left structures, indicating that the component clusters are also well-separated in high-dimensional space.    

\subsubsection{Cluster Properties}

In this section we discuss the clusters identified in run \med~(middle panels of Fig.~\ref{fig:umap_groups}) in more detail, using their characteristics as a proxy for the general core behavior across all simulations. Figure \ref{fig:umap_props} shows nine of the \med~core properties projected onto the UMAP. Figures for the other two runs are given in the appendix for comparison.

Inspection of the properties indicates that the radius of coherence, $R_{\rm coh}$, and outflow fraction drive the high-level organization in the UMAP. The UMAP structures on the left contain cores with a significant amount of coherence, while the rightmost structure has few cores with any amount of coherence. Similarly, cores that have the most amount of outflow material, namely those cores that are near  a protostar or host a protostar with an outflow, are sorted into the bottom half of the UMAP. Due to the clustered nature of star formation and outflow-driven turbulence \citep[e.g.,][]{OffnerChaban2017a,NeralwarColombo2024a}, a significant number of cores contain outflow material without having formed a protostar. Cores in the top-left structure have very little outflow material, and these small fractions are generally produced by turbulent mixing of outflow material rather than by close proximity to any outflows. These are the ``pristine" starless cores, removed from star formation activity. 

The rest of the variables create more subtle organization in the UMAP. There are strong north-south gradients in mass, size, velocity dispersion and magnetic energy, while trends with virial parameter are weaker.  Core shape creates east-west organization, with the most filamentary cores located on the bottom left and top right. Comparison with Fig.~\ref{fig:umap_groups} indicates that protostellar cores coincide with regions where the core shape is more spherical, which is consistent with gravitational influence.  Likewise, cores with high peak densities, found at the outer edge of the bottom left and rightmost structures, which is also a signpost of collapse, coincide with the locations of protostellar cores. 

Cores located in the bridge connecting the top and bottom leftmost large structures also exhibit property organization, mostly as an extension of the properties of the lower structure. Aside from the coherence radius and outflow fraction, the ranges of parameters in each of these two structures appear visually similar.

 The UMAP in Fig.~\ref{fig:umap_props} shows that cores containing sizable regions of coherence, which are organized into the outer top and bottom edges of the leftmost structures, do not necessarily have low overall velocity dispersions. This occurs because the velocity dispersion represents an average over all core gas, whereas the coherent region is localized to the core center, as defined by the density peak. The cores with the largest regions of coherence (up to 0.2 pc, which is the outer scale of the velocity dispersion profile) also tend to occur within the largest and most massive cores. While the overall core velocity dispersion is driven by turbulence, flows associated with core formation/accretion, and stellar feedback, cores can contain a quiescent central region where the turbulence has decayed and the velocity dispersion is lower. Comparison of Figures \ref{fig:umap_groups} and \ref{fig:umap_props} indicates that many starless cores are influenced by outflow material from nearby protostars, which interacts most directly with the outer envelopes of cores. 

Figure \ref{fig:groups} shows violin plots of the parameter distribution for each of the ten \med\ clusters, where the clusters are ordered by the portion of the UMAP they inhabit. The left half of each violin shows the properties for all cores, while the right half shows the distributions for protostellar cores. The top left panel includes a short-hand description of each structure: the top-left structure contains cores that are  mostly {\it starless, coherent and pristine}, the bridge indicates a {\it transition} region with rapidly changing properties, the bottom-left structure contains  many cores that have {\it coherent}  centers and are either {\it protostellar} or are interacting with protostellar outflows, and the right structure contains cores with {\it turbulent} centers and significant {\it protostellar} activity.

While the clusters have significant overlap in property range, clear quantitative differences are apparent. For example,  many of the largest, most massive and turbulent cores are grouped in cluster 5, which is located at the top of the top-left structure in the UMAP.  Cluster 6 also contains large, massive cores with high levels of turbulence, but these cores are less bound and exhibit significant outflow influence; they are located far from the cluster 5 cores in the bottom left of the UMAP. Cluster 3, which is the bridge between the upper and lower UMAP structures, and cluster 9 have distinctly large ranges in mass, radius, velocity dispersion and virial parameter that produce long, narrow violin plots. These two clusters are also extended in the UMAP, indicating that cores in these clusters span a broad range of properties. These two otherwise similar groups are primarily separated by the degree of core coherence. Cluster 9 is located in the right UMAP structure, together with non-coherent cores.

\subsubsection{Protostellar Core Cluster Membership}

Although the protostellar flag is not contained in the core property vector that is used for clustering, the protostellar cores are fairly neatly sorted into particular clusters. Clusters 1, 2 and 4, contain the large majority of protostellar cores. If a core is protostellar it has an 87\% likelihood of belonging to one of these three clusters. 

Cores in these clusters are characterized by high peak densities, low virial parameters and high outflow fractions, properties that are independently correlated with star-formation activity.  Cluster 4, in the bottom right of the UMAP, has the highest incidence of protostellar cores. A protostellar core has a 47\% likelihood of belonging to cluster 4, whereas there is a 20\% chance a protostellar core belonging to either clusters 1 and 2. 

Cores in cluster 2 have no region of coherence, indicating that some combination of gravitational collapse, accretion, and feedback are enhancing the internal velocity dispersion. Meanwhile,  cores in clusters 1 and 4 tend to have some degree of coherence. 


\section{Discussion}\label{sec:discuss}

\subsection{Comparison to Prior Numerical Models}

This study differs from prior numerical studies of dense cores in several important ways. The \starforge\ simulations provide a core catalog that is more than an order of magnitude larger than any core study to date. In addition, unlike prior work \citep[e.g.,][]{SmithClark2009a,OffnerKrumholz2008a,GongOstriker2015a, KuznetsovaHartmann2020a,SmullenKratter2020a,PelkonenPadoan2021a,OffnerTaylor2022a,CollinsLe2023a} all key feedback processes are included, allowing self-consistent dispersal of dense gas via outflows and radiative feedback. The inclusion of feedback also means that core environments vary significantly from the pristine conditions at the beginning of the simulations to the end when feedback from massive stars dominates the cloud energetics \citep{NeralwarColombo2024a}. Nonetheless, insofar as gravity and turbulence are key components of core formation and evolution, we find some commonalities with past work, which was based on simpler simulations.

Turbulence plays a key role in both creating structure, which leads to star formation, and dispersing dense gas, thereby reducing local gravitational collapse. By analyzing all over-densities, not only those that are self-gravitating  \citep{SmithClark2009a,PelkonenPadoan2021a}, we capture dense gas behavior that is not associated with star formation activity. 
As in \citet{OffnerTaylor2022a}, who adopt a similar core definition, we find that the large majority of the cores disperse before forming stars, e.g., only 6-9\% of well-behaved paths in these runs form protostars. In both studies, $<10$\% of all cores are protostellar. This occurs because most over-densities that reach peak densities above $10^4$\,cm\eee~with mass $>0.1\msun$ via turbulent compression, do not ultimately become self-gravitating. Even though the \starforge\ clouds globally collapse as the initial turbulence decays, in contrast to the periodic box simulation of \citet{OffnerTaylor2022a} where turbulence is produced by stochastic driving, here stellar feedback is sufficient to maintain local turbulence and significantly suppress collapse. 

The evolution and lifetimes of our prestellar cores are similar to those of cores in \citet{CollinsLe2024a}, who model a magnetized turbulent cloud with periodic boundary conditions and use tracer particles to track star-forming gas. 
Like the \starforge\ cores, those prestellar cores have a broad range of lifetimes and spend a significant amount of their evolution, $\sim$ 0.3-1 Myr, at intermediate densities. Likewise, the cores eventually collapse rapidly over $\sim 0.1$ Myr, a process which \citet{CollinsLe2024a} term ``cruise and collapse." In both sets of simulations the prestellar phase turbulence slowly decays, causing a gradual reduction in the core velocity dispersion. This evolution appears to be a general characteristic of dense gas in turbulent flows within weakly self-gravitating clouds \citep[see also][]{GongOstriker2011a,MoonOstriker2024a}.


\subsection{Observational Prospective and Context}

While we do not produce synthetic observations, which would allow apples-to-apples comparisons to dust continuum and molecular line surveys of nearby star-forming regions, we can compare to a number of properties and relations that are based on derived core properties.

\subsubsection{Observed Core Properties}

Our median core masses of $\sim 0.4-0.7\,\msun$ are comparable to the masses of dense cores in nearby regions. For example, a NH$_3$ survey of Ophiuchus, Taurus and Perseus by \citet{KerrKirk2019a} found median starless core masses of $0.4_{-0.3}^{+0.4}~\msun$. \citet{ChenPineda2019a} identified a sub-sample of 23 coherent cores in these regions, i.e., ``droplets," that are gravitationally unbound and have masses $0.2_{-0.1}^{+0.3}~\msun$. Meanwhile, a dust continuum survey of 237 starless and protostellar cores in Orion A using JCMT found $M_c = 0.8_{-0.4}^{+0.3}~\msun$\citep{KirkFriesen2017a}. As in the observations, high-mass starless and protostellar cores are relatively rare. The range in observed median masses may be partially due to variance in cloud conditions, including temperature and gravitational binding, but it is likely also influenced by how cores are identified, since all three of these surveys adopt different core definitions. 

The \starforge\ cores tend to be larger and less compact than their observational counterparts. The median \starforge\ core effective radius is $\sim 0.08$\,pc, while \citet{KirkFriesen2017a}, \citet{KerrKirk2019a}, and \citet{ChenPineda2019a} find median radii of $0.026_{-0.005}^{+0.01}$\,pc, $0.023_{-0.003}^{+0.008}$\,pc and $0.033_{-0.008}^{+0.01}$\,pc, respectively. Part of this difference may also be due to differences in the core definition and approach, since the observed radius is computed in projection and generally by assuming the cores are elliptical,  whereas we adopt a mass-weighted average as given in \S\ref{sec:construct_core}. 

The median \starforge\ core velocity dispersions of $\sim0.27-0.42$\,km\,s\e~ or $\sim1-2\,c_s$ are comparable to those of observed cores. \citet{KirkFriesen2017a}, \citet{KerrKirk2019a}, and \citet{ChenPineda2019a} find median velocity dispersions of $0.32_{-0.04}^{+0.02}$\,km\,s\e, $0.37_{-0.05}^{+0.09}$\,km\,s\e~ and $0.23_{-0.02}^{+0.01}$\,km\,s\e, respectively. A large fraction of the simulated cores are not only sub-sonic but have significant regions of coherence like the droplets studied in \citet{ChenPineda2019a}. Given the typical observational survey resolution, where small cores span only 1-2 beams, we expect significantly more cores to have coherent regions than currently reported.

Similar to observations, we adopt a simple definition for the virial ratio rather than conduct a full virial analysis.  While the range of core virial parameters appears large, $\alpha_{\rm vir}\sim0.1-100$, this span is consistent with the significant spread in observed cloud and core virial ratios \citep{KauffmannPillai2013a}. For example, the Orion A cores have $\alpha_{\rm vir} \sim 0.4-900$ \citep{KerrKirk2019a}. While relatively few of these are gravitationally bound, most appear to be pressure-confined by the weight of the cloud like many observed cores \citep[e.g.,][]{Maruta_2010,Pattle_2015,KerrKirk2019a}. As in our simulations, pressure-bound cores may not be as transient as they seem given their low binding energies, even if they are not ultimately star-forming.  
We also note that \citet{SinghMatzner2021a} demonstrate that observational estimates of $\alpha_{\rm vir}$ are subject to systematic errors related to background/foreground subtraction, which act to systematically reduce the virial ratio. This point also underscores that many observed structures are less bound than they appear and thus more likely to eventually disperse. 

While there are quantitative differences in the observed and simulation core property distributions, the qualitative similarities give confidence that the \starforge\ cores are good representations of the cores detected in nearby star-forming regions.

\subsubsection{Scaling Relations}

Historically, dense cores and clouds have been found to obey a set of scaling relations. \citet{Larson_1981} first showed that molecular clouds follow a `linewidth-size' relation, where the velocity dispersion scales as the structure size following $\sigma \propto R^{p_v}$, where $p_v = 0.38$ in the original empirical fit. This trend has been born out by a variety of later observational studies and attributed to turbulence, where supersonic turbulence corresponds to  $p_v = 0.5$ \citep{SolomonRivolo1987a,Goodman_1998,KauffmannPillai2013a}. 
However, the relation is not universal. Massive cores do not seem to obey the linewidth size relation \citep{PlumeJaffe1997a,BarnesYonekura2011a}, nor do the dense cores identified in the Greenbank Ammonia survey, which even exhibit a negative correlation where $p_v = -0.34$ \citep{KirkFriesen2017a,KerrKirk2019a,ChenPineda2019a}. This suggests that some cores are already decoupled from the turbulence cascade and another process dictates their internal velocity dispersion. 

Here, starless cores have a linewidth-size relation with $p_v \sim 0.26-0.36$, which suggests they are mostly turbulently regulated. In contrast, protostellar cores exhibit a slightly negative correlation with $p_v \sim -0.02-0.05$. Their higher overall velocity dispersions, compared to starless cores, indicate that stellar feedback rather than turbulence inherited from the cloud environment drives their internal motions and, thus, it is unsurprising that they do not exhibit a positive linewidth trend with size.

\citet{Larson_1981} also found that molecular structures follow a mass-size relation, $M \propto R^{p_m}$, where $p_m = 1.9$. A scaling of $p_m = 3$ corresponds to cores with fixed, constant density, while $p_m = 2$ suggests that structures have a fixed, constant surface density. A flatter scaling with $p_m = 1$  indicates some correlation between density and size, i.e., $\rho \propto R^{-2}$, where smaller cores are centrally condensed and larger structures are effectively fluffier. Here, we find that starless cores have $p_m = 1-1.2$, while protostellar cores exhibit even flatter scaling with $p_m\sim 0.7-1.0$. This is consistent with smaller structures being more centrally condensed. Indeed, our protostellar cores are systematically more compact. However, we note that inferred trends are sensitive to both the core definition and observational tracer, which can lead an order unity change in $p_m$ \citep{KerrKirk2019a,OffnerTaylor2022a}.

\subsubsection{Timescales}

Core lifetimes are a topic of longstanding debate in the star-formation community. Theoretical models for core lifetimes span two extreme paradigms: slow quasi-static contraction regulated by magnetic and turbulent support \citep{MouschoviasCiolek1999a,KrumholzTan2007b} or rapid turbulence-induced collapse dominated by gravity \citep{Ballesteros-ParedesKlessen2003a}. These models bracket predicted core lifetimes of $\sim$1-10 freefall times. Observational estimates often infer intermediate lifetimes of a few freefall times. For example, wide-field surveys of cores infer intermediate lifetimes of 2-5 freefall times or $\sim$1 Myr \citep{WardThompson_2007,Andre_2010}. However, these observational estimates are derived statistically, extrapolating from an assumed T Tauri lifetime of 1 Myr, which is itself highly uncertain. 

There is some indication that the lifetime scales with the core mean density such that lower density cores live for more freefall times \citep{WardThompson_2007}. For example, 
\citet{DasBasu2021a} inferred
lifetimes that range from 6 freefall times for cores with densities of $n=10^{4}$\,cm\eee\ to $\sim$1 freefall time for cores with densities of $n=10^6$\,cm\eee. However, these estimates were derived from model fits that a priori assume cores are well-described by a magnetically dominated flow, where the magnetic flux is mediated by ambipolar diffusion.

Studies based on the interferometic detection rate of a compact core center provide another independent probe of core properties, fragmentation, and lifetime \citep{OffnerMoe2023a}. This ``substructure" within cores can only be detected once central densities exceed $> 10^7$ cm$^{-3}$ \citep{DunhamOffner2016a}. Consequently, substructure detection is a signpost of incipient collapse.

Studies of substructure incidence combined with synthetic observations of star-formation simulations suggest that collapse, once it begins, must be fast, since substructure is rarely detected \citep{DunhamOffner2016a,KirkDunham2017a,FielderKirk2024a}. The frequency of substructure is best reproduced by a slow model of core collapse that is magnetically and turbulently regulated, i.e., conditions similar to those modeled in \starforge, rather than smooth, non-turbulent conditions, which undergo rapid freefall. Collapse in the latter case happens so quickly that no core substructure should be detected in current interferometric continuum surveys given the statistical sample size. However, the measured occurrence rate is complicated by the fact that, as in our simulations, many observed cores are pressure-confined and/or transient and may never collapse \citep{DunhamOffner2016a,OffnerTaylor2022a}.

Estimating core lifetimes from gas tracers that function as ``chemical clocks" is another promising approach that is orthogonal to those above.  A small study of deuterium fractionization in high-mass cores combined with chemical modeling suggests a relatively long lifetime of 10 freefall times \citep{KongTan2016a}. Meanwhile, low-mass cores in Taurus exhibit significant variation in the depletion of CO and N$_2$H$^+$, indicating a range of formation and evolution pathways spanning slow to rapid collapse \citep{LeeEvans2003a,ChoiLee2017a}. However, these inferred evolutionary times depend on accurately estimating the underlying gas conditions and are sensitive to the size of the model chemical network, where small ``reduced" networks can lead to factors of two error in the estimated lifetime \citep{SipilaCaselli2022a}.

In the present study, the \starforge\ protostellar core lifetimes are comparable to several freefall times (see 
\S\ref{sec:paths}). The median starless core lifetimes are only 1-2 freefall times given a mean core density of a few  $10^{4}$\,cm\eee. However, as in the case of observations, the core properties and histories are very heterogeneous, with a subset of cores undergoing rapid collapse (e.g., left panels of Fig.~\ref{fig:coreevo}), while other starless and protostellar cores live for more than 10 freefall times as shown in Fig.~\ref{fig:life}. 

While we expect our simulated cores to be detectable in continuum and molecular surveys (e.g., in NH$_3$ and N$_2$H$^+$) given their peak densities \citep{FriesenPineda2017a,BettiGutermuth2021a,PriestleyClark2023a}, we stress that cores are identified in observations and simulations quite differently, which makes direct comparison challenging. In addition, many of the theoretical models underlying the studies above assume isolated cores and neglect the formation and accretion of gas \citep[e.g.,][]{MouschoviasCiolek1999a,DunhamOffner2016a,SipilaCaselli2022a}, which would lengthen the core lifetime. 

The median protostellar core lifetime of $\sim 0.1$ Myr (Table \ref{tab:paths}) is comparable to the observed protostellar Class 0 lifetime, which represents the most heavily embedded stage of star formation. Estimates derived by counting the number of protostars in different classes suggest a Class 0 lifetime of $\sim 0.15$ Myr and a combined Class 0/I lifetime of $\sim 0.5$ Myr \citep{DunhamStutz2014a}. Statistical arguments that do not assume that the star formation rate is constant infer shorter Class 0 and I lifetimes of 0.047 Myr and 0.088 Myr, respectively \citep{KristensenDunham2018a}. However, both these approaches calibrate the lifetimes using an assumed Class II/disk lifetime of 2 Myr, which is highly uncertain and likely varies from region to region.



\subsection{Numerical Approximations and Caveats}

While the simulations include all major feedback mechanisms and physical processes, a few simplifications remain. The initial conditions, as in most numerical star-formation studies, assume the molecular cloud is isolated from the larger galactic environment, and thus, it does not accrete material from its surroundings. Although the median core lifetime is more than an order of magnitude smaller than the cloud lifetime, 
the longest core lifetimes are bookended by the initial conditions, which determine the formation time and evolution of the first cores before turbulence is well-developed.  Simulations including self-consistent cloud formation and interaction with the larger galactic environment are required to determine the impact of ongoing accretion and cloud evolution on the dense gas \citep[e.g.,][]{HopkinsGrudic2024c}.

These \starforge\ simulations also neglect non-ideal MHD processes such as ambipolar diffusion, Hall effect and ohmic dissipation. These processes regulate the gas-magnetic field coupling and angular momentum transport, which may impact the evolution of core properties \citep{ZhaoTomida2020a}. Non-ideal effects may also reduce the maximum magnetic field strength in the cores, reducing the overall magnetic support and leading to earlier collapse \citep{WursterBate2021a}.

Finally, the simulations neglect self-consistent cosmic-ray (CR) transport and feedback, i.e., CRs accelerated in accretion, outflow, and wind shocks \citep{PadovaniIvlev2020a}. While CRs mediate the temperature and chemistry of the dense gas, there appears to be relatively little effect on cloud temperatures and evolution for typical Milky Way CR conditions when self-consistent CR transport is included \citep{FitzAxenOffner2024a}. CRs accelerated locally may provide additional heating and ionization \citep{GachesOffner2018b,GachesOffner2019a,Fitz-AxenOffner2021a,PinedaSipila2024a}, but this likely has a relatively small impact on the core dynamics. 

\section{Conclusions}\label{sec:conclude}

This work presents the analysis of the largest statistical sample of dense cores to date.  We identify more than 1,000,000 cores across three \starforge\ simulations with varying magnetic field strengths. We track the core evolution from birth through dispersal, identifying more than 200,000 unique paths. Since the calculations include all key physics and stellar feedback processes and end when the parent molecular cloud disperses, we are able to self-consistently examine the impact of environment and stellar feedback on core evolution.  

We find that cores undergo many merges and splits over their lifetime; consequently we focus the analysis on the subset of cores in well-behaved paths, which are defined to be either starless throughout or those that become protostellar and remain so until dispersal. We find that the large majority of cores disperse before forming stars, such that only 6-8\% of cores in well-behaved paths are protostellar. The median starless core lifetime is $\sim 0.5-0.6$\,Myr and the median protostellar core lifetime is $\sim$0.8-1.1\,Myr; however, core trajectories span a broad range of lifetimes, from one to more than 10 freefall times, with no distinct characteristic timescale. In all three runs the protostellar phase lasts $\sim 0.1^{+0.1}_{-0.05}$\,Myr, at which point feedback disperses the gas and the core mass falls below the detection limit of $0.1\msun$.

As in previous works that consider all over-densities \citep{SmullenKratter2020a,OffnerTaylor2022a}, we find that core evolution is very stochastic, partially due to the nature of turbulent flow and partially due to the lack of a well-defined core boundary (as identified using dendrograms). However, some trends are apparent: star-forming cores experience a gradual decline in their kinetic and magnetic virial ratios during the starless phase, coincident with turbulent decay. Once turbulence declines and the central density exceeds $\gtrsim 10^6$cm\eee, runaway collapse occurs.

We find weak trends in median mass, radius, and velocity dispersion with cloud magnetic field. Cores forming in the weaker field clouds have smaller masses and radii and larger velocity dispersions. The median core kinetic and magnetic ratios also decrease with increasing cloud magnetic field such that the identified cores 
are more self-gravitating. The cores display relatively flat linewidth-size and mass-size relations with $\sigma \propto R^{0.3}$ and $M \propto R^1$, where protostellar cores show no correlation between linewidth and size, indicating that their dispersions are regulated by stellar feedback not the cloud turbulent cascade.  Our core properties, including mass, size, linewidth and virial parameter, are in good agreement with those derived for cores identified in NH$_3$ in local star-forming regions \citep{KirkFriesen2017a,KerrKirk2019a,ChenPineda2019a}.

For each core, we construct a vector of 25 properties, apply a clustering algorithm, and project the result to a 2-D space using UMAP. We identify 6-10 distinct groups, where protostellar cores have a $>80$\% likelihood to be found in three particular groups, which are characterized by high central densities, compact radii, and lower virial parameters. Given the stochastic nature of core evolution, the cores are not significantly more likely to form or disperse from any one group. In all runs, we find that the identified groups are mapped to three major structures in the UMAP, similar to those in \citet{OffnerTaylor2022a}, which are separated by the degree of outflow feedback present in the core and whether the core is coherent. The groups exhibit a high degree of organization in the UMAP by mass, size, velocity dispersion, magnetic field strength, shape and peak density. However, aside from high-peak density, which is a signature of collapse, no particular set of properties is predictive of the core outcome.

We find some monotonic trends in core properties with environment (magnetic field strength), consequently we expect additional differences to occur under more extreme cloud conditions. For example, further investigation is needed into the impact of varying global virial parameter, metallicity, and column density on core formation and evolution. Finally, while our calculations are reasonably complete in terms of the included physics and feedback, future work is required to explore dense cores in models with self-consistent cloud formation, which may impact the early core histories and lifetimes.

\vspace{0.1in}
The authors thank Brandt Gaches, Kartik Neralwar, and an anonymous referee for helpful comments.
SO and JT were supported by NSF AST-2107942. SO also acknowledges support from NSF CAREER 1748571, NSF AST-1812747, NASA 80NSSC23K047, and NSF under Cooperative Agreement 2421782 and the Simons Foundation award MPS-AI-00010515.
This research made use of Astropy, a community-developed core PYTHON package for Astronomy \citep{astropy}.

\bibliography{main,ml,biblio_complete_2023}{}
\bibliographystyle{aasjournal}

\appendix

\begin{figure*}
    \centering
    \includegraphics[width = 1.0\textwidth, keepaspectratio]{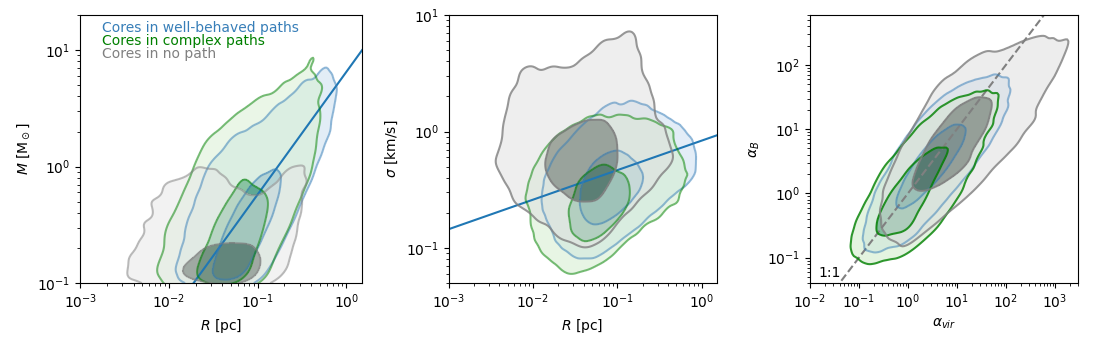}
    \includegraphics[width = .95\textwidth, keepaspectratio]{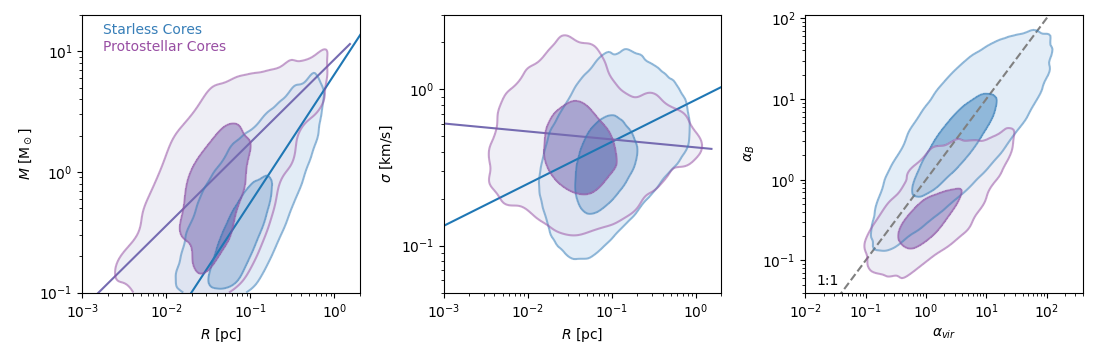}
 \caption{{\bf Appendix:} Distribution of core properties for simulation \lo. Top row: Cores in well-behaved paths are blue, cores in paths that behave non-monotonically are in green, and identified cores that are unconnected to any path are in grey. Bottom row: Starless cores in well-behaved paths are in blue and protostellar cores in well-behaved paths are purple.  Left: mass versus radius. Middle: Velocity dispersion versus radius. Right: Magnetic virial parameter versus virial parameter.}
    \label{fig:coreprop_lo}
    \end{figure*}

\begin{figure*}
    \centering
    \includegraphics[width = 1.0\textwidth, keepaspectratio]{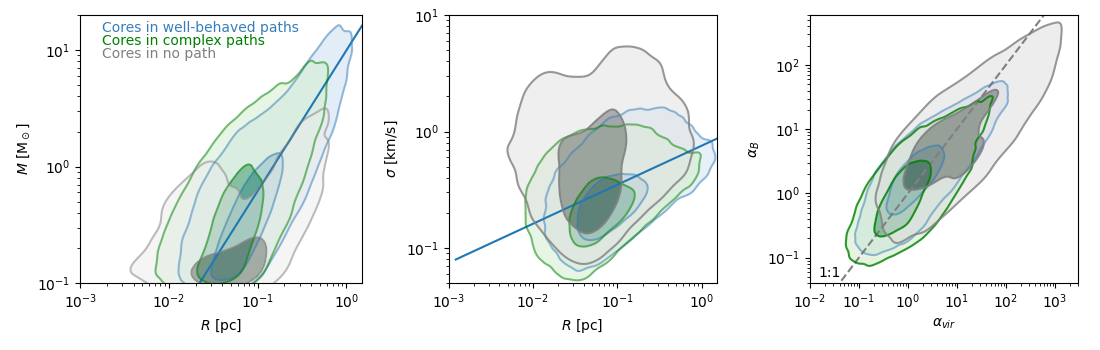}
    \includegraphics[width = .95\textwidth, keepaspectratio]{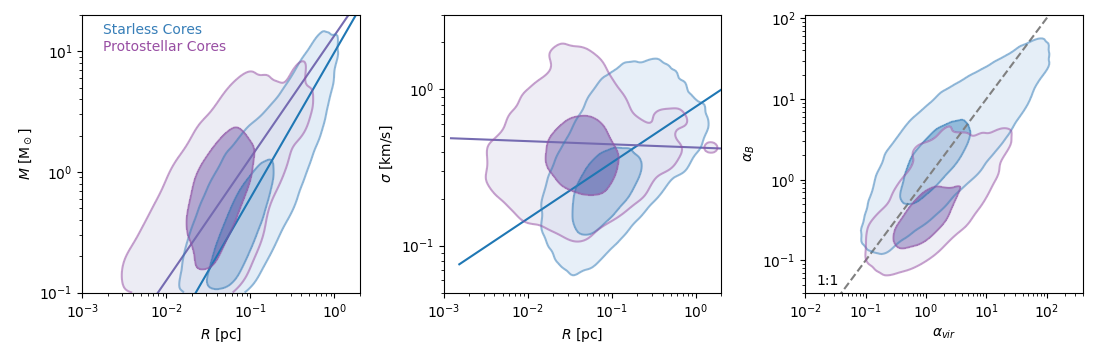}
 \caption{{\bf Appendix:} Distribution of core properties for simulation \hi. Top row: Cores in well-behaved paths are blue, cores in paths that behave non-monotonically are in green, and identified cores that are unconnected to any path are in grey. Bottom row: Starless cores in well-behaved paths are in blue and protostellar cores in well-behaved paths are purple.  Left: mass versus radius. Middle: Velocity dispersion versus radius. Right: Magnetic virial parameter versus virial parameter.}
    \label{fig:coreprop_hi}
    \end{figure*}

Figures \ref{fig:coreprop_lo} and \ref{fig:coreprop_hi} show the distributions of core properties for the  \lo\ and \hi\ simulations, respectively. The distributions, trends, and fits are qualitatively similar to those of the fiducial run shown in Fig.~\ref{fig:coreprop}.

\begin{figure*}
    \centering
    \includegraphics[width = .95\textwidth, keepaspectratio]{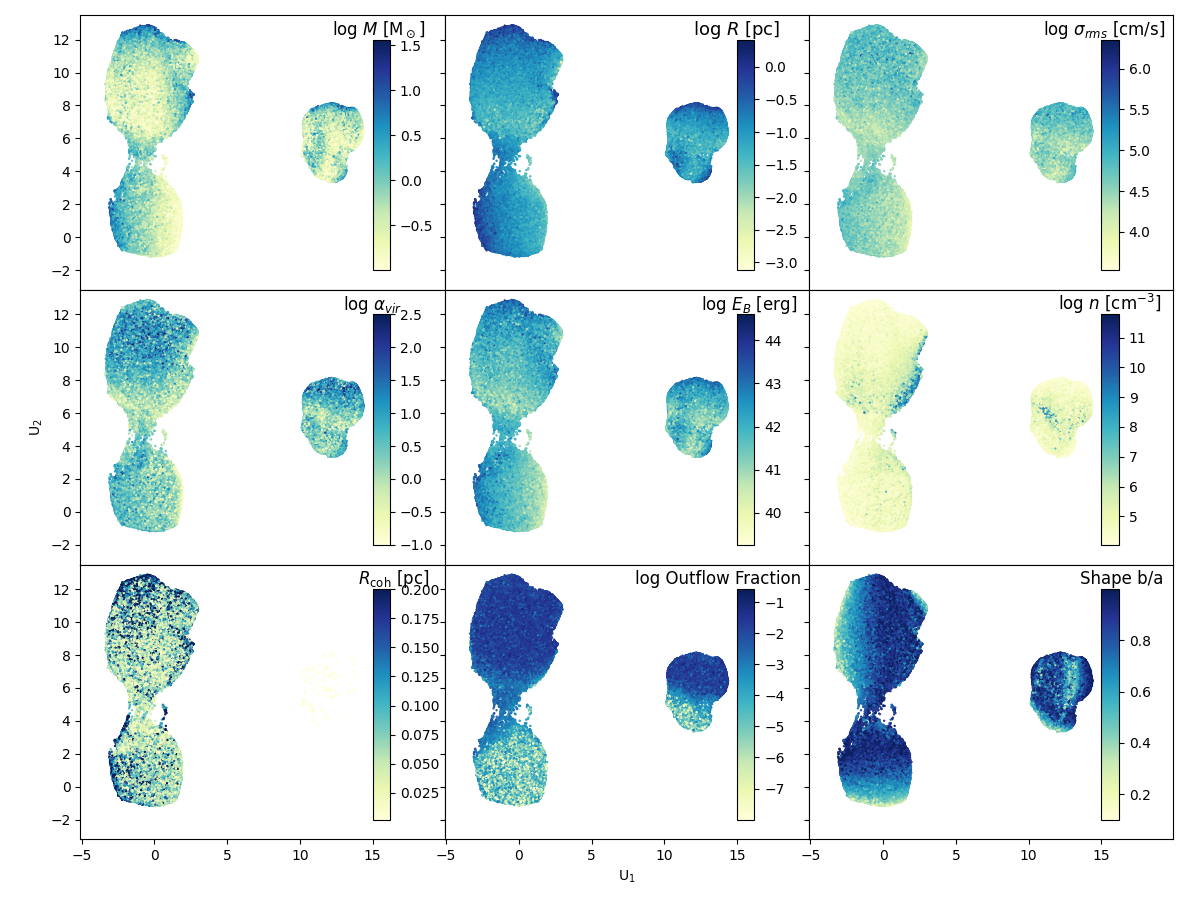}
 \caption{ {\bf Appendix.}  UMAP projection of \lo core properties, where each dot represents a core belonging to a well-behaved path. Top: UMAP colored by core mass, radius, and velocity dispersion from left to right. Middle: UMAP colored by virial parameter, magnetic energy and peak density from left to right. Bottom: UMAP colored by the size of the region of coherence ($R_{\rm coh}$ defined as the outer radius where $\sigma_{\rm rms}(r) < c_s$), mean outflow mass fraction, and aspect ratio.  The shape of the projected distribution is strongly influenced by whether cores are coherent (left structures) and whether they contain a significant fraction of outflow material (lower structures). The bulk properties shape the ordering of the cores within each region.}
    \label{fig:umap_props_lo}
    \end{figure*}

\begin{figure*}
    \centering
    \includegraphics[width = .95\textwidth, keepaspectratio]{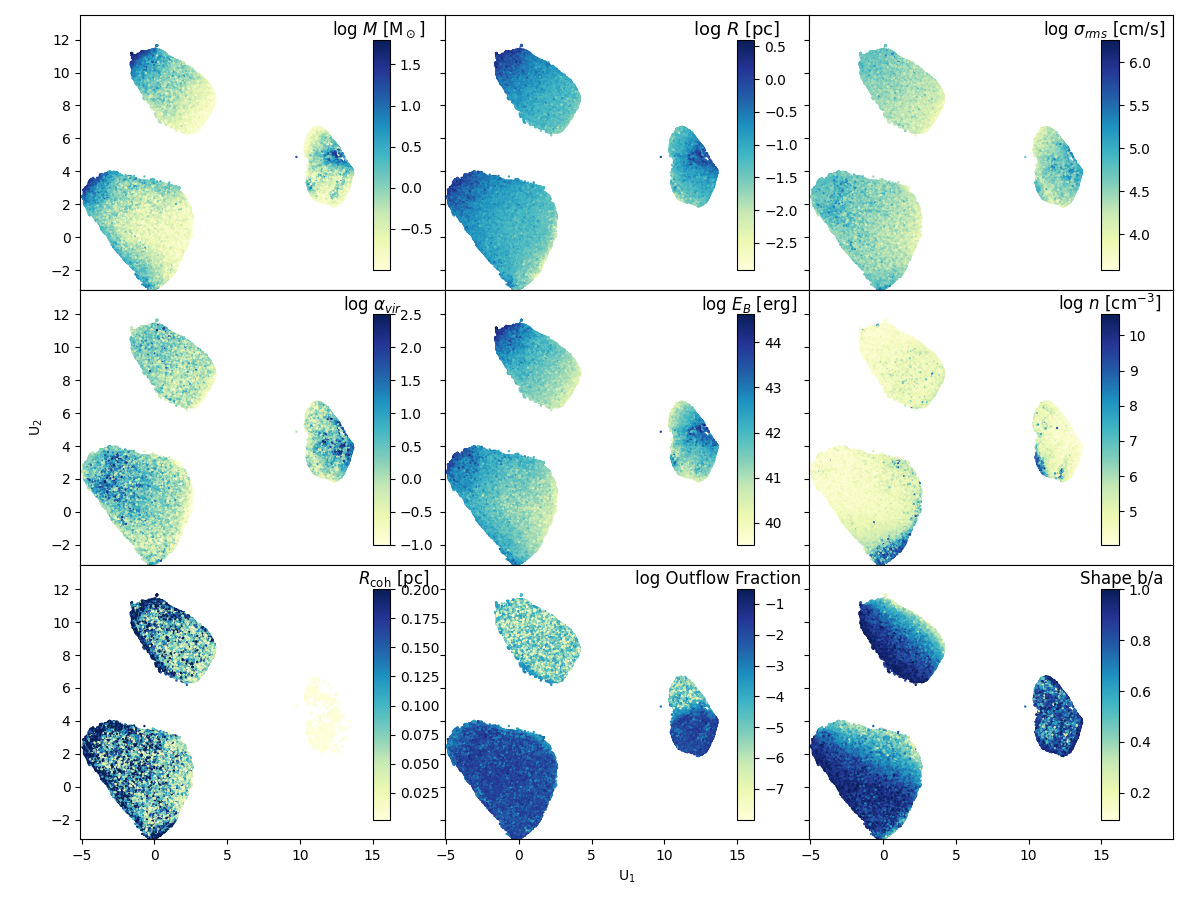}
 \caption{{\bf Appendix.} UMAP projection of \hi core properties, where each dot represents a core belonging to a well-behaved path. Top: UMAP colored by core mass, radius, and velocity dispersion from left to right. Middle: UMAP colored by virial parameter, magnetic energy and peak density from left to right. Bottom: UMAP colored by the size of the region of coherence ($R_{\rm coh}$ defined as the outer radius where $\sigma_{\rm rms}(r) < c_s$), mean outflow mass fraction, and aspect ratio.  The shape of the projected distribution is strongly influenced by whether cores are coherent (left structures) and whether they contain a significant fraction of outflow material (lower structures). The bulk properties shape the ordering of the cores within each region.}
    \label{fig:umap_props_hi}
    \end{figure*}

Figures \ref{fig:umap_props_lo} and \ref{fig:umap_props_hi} show the UMAPs of the core properties for the simulations \lo\ and \hi, respectively. These properties show a similar level of structure separation by core coherence and outflow fraction. Each of the larger structures also shows a strong degree of ordering in mass, size, velocity dispersion, magnetic energy and shape.

\end{document}